\documentstyle[epsf]{mn}

\title{Long-term IR Photometry of Seyferts}

\author[I.S. Glass]
{I.S. Glass\\
South  African   Astronomical Observatory, PO Box 9, Observatory 7935, 
South Africa}

\date{Received 2003}

\begin{document}

\maketitle

\begin{abstract}

Long-term (up to 10000d) monitoring has been undertaken for 41 Seyferts
in the near infrared (1.25 -- 3.45$\mu$m). All but two showed variability,
with amplitudes at $K$ in the range $<$ 0.1 to $>$1.1 mag. The timescale for
detectable change is from about one week to a few years.

Where contemporary observations of variability in x-rays, UV or visible
light exist, it is found that the near-infrared varies in a similar way,
though in some cases the shorter-wavelength IR bands are diluted by
underlying galaxy radiation.

A simple cross-correlation study indicates that there is evidence for delays
of up to several hundred days between the variations seen at the shortest
wavelengths ($U$ or $J$) and the longest ($L$) in many galaxies. In
particular, the data for Fairall 9 now extend to twice the interval covered
in earlier publications and the delay between its UV and IR outputs is seen
to persist.

An analysis of the fluxes shows that, for any given galaxy, the colours of
the variable component of its nucleus are usually independent of the level
of activity. The state of activity of the galaxy can be parameterized. 

Taken over the whole sample, the colours of the variable components fall
within moderately narrow ranges. In particular, the $H-K$ colour is
appropriate to a black body of temperature 1600K. The $H-K$ excess for a
heavily reddened nucleus can be determined and used to find $E_{B-V}$, which
can be compared to the values found from the visible region broad line
ratios.

Using flux-flux diagrams, the flux within the aperture from the underlying
galaxies can often be determined without the need for model surface
brightness profiles. In many galaxies it is apparent that there must be an
additional constant contribution from warm dust.

\end{abstract}   

  
\section{INTRODUCTION}

This programme was started with the aim of making a reliable study of the
variability of Seyfert galaxies in the near infrared ($JHKL$ bands; 1.25,
1.65, 2.2 and 3.45$\mu$m). Previous work had indicated that some Seyfert
galaxies do vary, but the data remained somewhat sparse and controversial.
At the commencement of the programme it was believed, for example, that
Seyfert 1 galaxies varied but Seyfert 2 galaxies did not. A 450-day
programme by Penston et al (1974) appeared to confirm this viewpoint.
Lebofsky \& Rieke (1980) reported on variations observed in several
Seyferts, of which that in 3C\,120 was the most spectacular, but the numbers
of observations were very limited.

Some results from this programme have already been published. One of the
most unexpected was the discovery by Clavel, Wamsteker and Glass (1989) that
there was a delay of ~400 days between the UV variations of Fairall 9 and
the response of its infrared output. This was satisfactorily interpreted
according to a dust reverberation model (Barvainis, 1987). NGC\,1566 showed
a delay of 2 $\pm$ 1 month between the UV and the IR (Baribaud et al, 1992).
Similarly, a possible delay of $\sim85$d was seen in NGC\,3783 (Glass,
1992). The infrared output of NGC\,1068 was shown to have increased by a
factor of two between the early 1970s and 1995 by Glass (1995). The galaxy
NGC\,2992 showed an energetic flare-like outburst with a decay time of about
900d (Glass, 1997a) and NGC\,7469 showed a dramatic but relatively
short-lived decline in its nuclear energy output in October 1989 (Glass
1998). 

More recently, Salvati et al (1993) observed an outburst in NGC\,4051 and
Nelson (1996) reported one in Mkn\,744, with a 32 $\pm$ 7 day delay between
the visible and IR light curves. Oknyanskij \& Horne (2001) have re-examined
the published data and concluded that the observed delays are consistent
with dust sublimation radii $R \propto \sqrt{L_{\rm UV}}$. Oknyanskij (2002)
includes two additional objects. The angular diameter of the dust ring or
shell as viewed fron the earth is of order tenths of milliarcsec.

Throughout this paper it is assumed that the excess infrared radiation from
a Seyfert when compared to a less active galaxy comes from dust that is
heated by ultraviolet radiation originating near a central massive black
hole. It is usually assumed that this dust is distributed in a torus; e.g.,
Pier \& Krolik (1992). The maximum temperature that dust can reach without
sublimating, in the neighbourhood of 1600K, together with the ultraviolet
luminosity of the nucleus, sets the inner radius of its distribution and the
corresponding light travel time gives the delay in the response of a dust
grain to a change in the ultraviolet flux from the nucleus. 

During part of the present programme, $UBVRI$ photometry was obtained by
Winkler et al (1992), Winkler (1997) and other observers.

\section{Observations}

The choice of galaxies to observe was very simple: only a few southern
Seyferts were known at the start of the programme. As others were
discovered, by examination of the ESO and UK Schmidt plates or from x-ray
satellites, they were included. The $L$-band luminosities (calculated from
the redshifts and the fluxes seen through a 12 arcsec diameter aperture)
range over three orders of magnitude. It will be seen later that the
proportion of nuclear to underlying galaxy fluxes also covers a large range.

Nearly all the observations were made with the MkIII infrared photometer
attached to the 1.9m telescope at Sutherland.  The aperture was normally 12
arcsec diameter.  In some cases, $L$ observations were made at 9 arcsec
diameter to improve the signal-to-noise ratio at slight expense to the
systematic accuracy. The chopper throw was 30 arcsec, alternately north and
south of the object position. The same filters were used throughout the
programme.

Some of the earliest observations were made with the MkI infrared
photometer, which used the same filters but had chopper throws of 12 or 60
arcsec, also in the N--S direction.

Standard stars were taken from Carter (1990), whose $J$ measures were
transformed to the natural system of the photometer using the empirical
relation $J_{\rm natural}=J_{\rm Carter} - 0.05(J-H)_{\rm Carter}$. The
other bands were taken to be the same on both systems. The results were
placed on the Carter system using the inverse transformation.

Table 1 lists the galaxies included in the survey. The positions, redshifts,
Seyfert types, $l$, and $b$ were taken from V\'{e}ron-Cetty \& V\'{e}ron
(2000). The absorption $E_{B-V}$ arising in our own galaxy was calculated
from the NASA/IPAC Database Extinction Calculator on the Worldwide Web,
which is based on Schlegel, Finkbeiner \& Davis (1998). The columns $\Delta
J$ and $\Delta K$ are the overall observed amplitudes observed at $J$ and
$K$, respectively. It should be noted that these two quantities are
sensitive to errors in individual measurements, and are, in effect, upper
limits. The column `No' gives the number of observations of each object. The
last three columns give the average dereddened near-IR colours.

Observations were usually made four times per year over periods of a week.
Usually a given galaxy could be observed three times per year, with one or
two repeats within a few days. In a few cases, several observations were
made over a short period as part of international multi-wavelength
campaigns.

Figs 2 a,b,c,d show the observations plotted against time. The same time and
magnitude scales are employed throughout, for ease of comparison. Additional
data obtained by other workers in the $U$ band have been included in several
cases. The error bars have been plotted only when they are expected to be
substantial. Normally, the standard errors at $JHK$, arising from the
standard stars and the uncertainties in the extinctions, are expected to be
${\stackrel{<}{\sim}}$0.03 mag and, at $L$,
${\stackrel{<}{\sim}}$0.05 mag. In general, these are the dominant errors.
However, at $L$ the signal-to-noise ratio was reduced and observations were
usually limited to about 320 sec of integration, so that the final errors
may be higher than 0.05 mag, especially when $L$ $>$ 9.

The observations are available at the end of this version of the paper.

\begin{table*}
\begin{minipage}{17.5cm}
\caption{List of galaxies in survey}
\begin{tabular}{ll@{\hspace{1mm}}l@{\hspace{0.5mm}}rl@{\hspace{1mm}}l@{\hspace{0.5mm}}llllllllllllll}

Name          &  \multicolumn{6}{c}{R.A.$^1$ (2000) Dec.}   &  z$^1$ & type$^1$&
$L_{3.4\mu{\rm m}}$ &
$E_{B-V}$& $\Delta J$ & $\Delta K$ & No. & $J$-$H$&$H$-$K$&$K$-$L$\\
III Zwicky 2  & 00&10&31.0& +10&58&28&.090&S1.2&23.72&.082&.66& .68&17&0.91&1.09&1.45\\ 
Tol 0109-38   & 01&11&27.7&--38&05&01&.011&S1.9&22.88&.014&.21& .34&31&1.27&1.14&1.76\\
F9            & 01&23&45.8&--58&48&21&.045&S1.2&23.66&.020&.78&1.15&71&1.00&0.91&1.45\\
NGC526A       & 01&23&54.4&--35&03&56&.019&S1.9&22.69&.022&.26&1.01&52&0.96&0.74&1.31\\
NGC985        & 02&34&37.8&--08&47&15&.043&S1.5&23.18&.030&.31& .48& 7&0.89&0.66&1.13\\
ESO 198-G24   & 02&38&19.7&--52&11&32&.045&S1.0&23.10&.033&.62& .92&18&0.90&0.74&1.23\\
NGC1068       & 02&42&40.7&--00&00&47&.003&S1h &22.88&.030&.12& .60&27&1.01&1.03&2.16\\
NGC1566       & 04&20&00.7&--54&56&17&.004&S1.5&21.37&.008&.14& .28&42&0.83&0.29&0.42\\
3C120         & 04&33&11.1& +05&21&15&.033&S1.5&23.31&.297&.50& .53&30&0.93&0.90&1.41\\
Akn 120       & 05&16&11.4&--00&09&00&.033&S1.0&23.44&.109&.40& .52&32&0.96&0.82&1.28\\ 
MCG-5-13-17$^2$&05&19&35.6&--32&39&30&.013&S1.5&22.12&.017&.15& .23&25&0.77&0.36&0.70\\
Pic A$^3$     & 05&19&44.3&--45&46&50&.034&S1.5&  -  &.043&.42& .33&14&1.00&0.96&    \\
NGC2110       & 05&52&11.4&--07&27&23&.020&S1i &22.99&.375&.09& .10&15&0.85&0.34&0.69\\
H0557-383     & 05&58&02.1&--38&20&05&.034&S1.2&23.61&.046&.35& .37&33&1.14&1.11&1.78\\           
F265          & 06&56&17.3&--65&33&48&.029&S1.2&22.74&.082&.20& .22&12&0.87&0.65&1.07\\
IRAS09149-6206& 09&16&09.5&--62&19&29&.057&S1  &24.32&.182&.24& .26& 7&1.08&1.09&1.52\\
NGC2992       & 09&45&42.0&--14&19&35&.008&S1.9&22.17&.060&.38& .65&52&0.98&0.52&0.84\\
MCG-5-23-16$^4$&09&47&40.2&--30&56&54&.008&S1.9&22.16&.108&.19& .37&25&0.86&0.51&1.17\\
NGC3783       & 11&39&01.8&--37&44&19&.009&S1.5&22.44&.119&.63& .90&95&0.86&0.72&1.31\\
H1143-182$^5$ & 11&45&40.6&--18&27&17&.033&S1.5&22.77&.039&.70& .66&14&0.85&0.70&1.09\\
I Zwicky 96   & 12&00&43.3&--20&50&01&.062&S1  &  -  &.051&.18& .19&13&0.78&0.44&   \\
NGC4593       & 12&39&39.4&--05&20&39&.009&S1.0&22.26&.025&.17& .40&41&0.88&0.52&0.94\\
NGC4748$^6$   & 12&53&12.4&--13&24&53&.014&S1n &21.95&.050&.16& .28&14&0.79&0.38&0.43\\
ESO 323-G77   & 13&06&26.6&--40&24&42&.015&S1.2&23.12&.101&.31& .49&27&1.09&0.84&1.22\\
MCG-6-30-15   & 13&35&53.4&--34&17&48&.008&S1.5&22.21&.062&.18& .18&25&0.88&0.66&1.25\\
IC4329A       & 13&49&19.3&--30&18&34&.016&S1.2&23.22&.059&.29& .46&38&0.97&0.87&1.48\\
NGC5506       & 14&13&14.8&--03&12&26&.007&S1i &22.71&.060&.37& .70&43&1.50&1.27&1.59\\
NGC5548       & 14&17&59.6& +25&08&13&.017&S1.5&22.81&.020&.22& .19&12&0.86&0.72&1.29\\
IRAS15091-2107& 15&11&59.8&--21&19&02&.044&S1n &23.36&.118&.26& .27&10&0.92&0.82&1.33\\
ESO 103-G35   & 18&38&20.3&--65&25&42&.013&S1.9&22.20&.076&.16& .09&25&0.81&0.37&1.32\\
F51           & 18&44&54.0&--62&21&53&.014&S1.5&22.65&.108&.29& .37&29&1.00&0.84&1.34\\
ESO 141-G55   & 19&21&14.3&--58&40&13&.037&S1.2&23.36&.111&.59& .71&34&0.94&0.89&1.35\\
NGC6814       & 19&42&40.7&--10&19&24&.006&S1.5&21.50&.183&.19& .32&24&0.83&0.29&0.40\\
Mkn 509       & 20&44&09.7&--10&43&24&.035&S1.5&23.43&.057&.65& .62&46&0.91&0.88&1.35\\
1H2107-097    & 21&09&09.8&--09&40&17&.027&S1.2&23.00&.233&.26& .33& 7&0.90&0.79&1.29\\
1H2129-624    & 21&36&23.2&--62&24&00&.059&S1.5&23.33&.037&.49& .72& 6&0.94&0.80&1.31\\
NGC7213       & 22&09&16.4&--47&10&01&.006&S3b &22.14&.015&.12& .18&26&0.81&0.36&0.68\\
NGC7469       & 23&03&15.6& +08&52&26&.017&S1.5&23.05&.069&.31& .63&32&0.96&0.67&1.12\\
MCG-2-58-22   & 23&04&43.5&--08&41&08&.048&S1.5&23.44&.042&.87&1.17&49&0.94&0.80&1.31\\
NGC7603       & 23&18&56.7& +00&14&38&.029&S1.5&23.09&.046&.40& .87&26&0.89&0.59&1.05\\

\end{tabular}

\noindent Notes: Positions, redshifts and Seyfert types taken from V\'{eron}-Cetty \&
V\'{e}ron (2000). Pic A and NGC\,7213 have broad Balmer lines, but also
exceptionally strong [OI] and many other typical liner characteristics. In
the blue, NGC\,2110 looks like a type 2, but near H$\alpha$ it has the
spectrum of a liner (Winkler, private communication).

The $E_{B-V}$ values are taken from the NASA/IPAC Database Extinction
Calculator on the World Wide Web, which is based on Schlegel et al (1998).

The average luminosity at $L$, seen throught the 12 arcsec diameter
aperture, is given in units of log WHz$^{-1}$ in the column marked
$L_{3.4\mu{\rm m}}$, taking $H_0$ as 65 km s$^{-1}$pc$^{-1}$.

$\Delta J$, $\Delta K$ are maximum amplitude ranges observed.

$J-H$ etc are the unweighted averaged de-reddened colours.

$^1$These columns are from V\'{e}ron-Cetty \& V\'{e}ron (2000).

$^2$Given by V\'{e}ron-Cetty \& V\'{e}ron (2000) as ESO362-G18

$^3$Given by V\'{e}ron-Cetty \& V\'{e}ron (2000) as PKS0518-45. 

$^4$Given by V\'{e}ron-Cetty \& V\'{e}ron (2000) as ESO434-G40.

$^5$Given by V\'{e}ron-Cetty \& V\'{e}ron (2000) as NPM1G-18.0386.  

$^6$Better known as IRAS1249-131

\end{minipage}
\end{table*}

\section{Average near-IR colours}

The average, unweighted, de-reddened colours of the galaxies in the sample
are given in the right-most columns of Table 1. De-reddening was carried out
according to the relations
\[
E_{B-V}:A_J:A_H:A_K:A_L::1:0.742:0.430:0.245:0.094
\]

The $J-H$ and $H-K$ colours of `inactive', or early-type galaxies without
obvious emission lines, on the same photometric system as that used here,
are typically around 0.78 and 0.22 (Glass, 1984). The $K-L$ colour (in the
absence of dust emission) has been taken to be 0.22, based on the
similarity of $H-K$ and $K-L$ colours for late-type stars (Glass, 1997c) 

A few of the sample, NGC\,1566, MCG-5-13-17, NGC\,2110, I\,Zw\,96,
NGC\,4748, NGC\,6814 and NGC\,7213, show $J-H$ and $H-K$ colours only
moderately redder than inactive galaxies. This can also be seen in the
vertical spacing of their light curves in Fig 2. When quiescent, the $L$
light curve of a galaxy such as NGC\,1566 is so close to the $K$ curve that
there is obviously no $L$ excess. NGC\,4748 and NGC\,6814 also revert to
inactive colours during minima. Their nuclear fluxes are assumed to be
swamped by the stellar fluxes of the underlying galaxies at $J$ and $H$, but
may become more conspicuous at longer wavelengths because their spectral
energy distributions rise and overcome the Rayleigh-Jeans tails of the
stellar fluxes.

$K-L$ excesses are seen at all times in MCG-5-13-17, NGC\,2110 and
NGC\,7213, indicating continuous nuclear activity. NGC\,7213 has positive
$U-B$, indicating that its nucleus is heavily reddened.

On the other hand, in the case of ESO 103-G35, the $L$ and $K$ light curves
are far apart, implying a considerable excess in spite of a lack of
variability. Since its $H-K$ colour is normal, the $L$ excess must arise
from a blackbody of about 800K or less. A positive $U-B$ implies that its
nucleus may be heavily reddened.

Most of the remainder of the sample show well-developed infrared excesses. 

\section{Variability Patterns}

The galaxies of the sample do not show variability on a night-to-night
basis, within the limits of the photometry. The only ones which may not have
varied at all are NGC\,2110 and ESO\,103-G35. Most show smooth behaviour on
a timescale of years, but some show shorter-term fluctuations. The most
conspicuous of the latter is NGC\,3783. An active galaxy is thus more likely
to be revealed by IR variability than by broad emission lines, which may be
hidden from view by extinction within the host galaxy.

As already mentioned, two very interesting cases of large amplitude
short-term events are seen in NGC\,2992, which seems to have experienced a
substantial outburst (Glass, 1997a) and NGC\,7469, whose nuclear source
virtually shut off for a short period (Glass, 1998).

Some galaxies show major changes over the long term. For example, the
archetype S2 galaxy NGC\,1068 doubled in $L$ flux over about 20 years.
NGC\,4748 has become fainter over 2000d and ESO\,323-G77 over 5000d.
MCG-2-58-22 has declined by over a magnitude in 5000d, though not
monotonically.
 
Otherwise, major variations, of more than one mag, have been seen in
Fairall 9 and NGC\,526a. Fairall 9 shows a delay between its UV output and
its $L$ flux, which can be detected even between the $J$ and $L$ bands. This
has been satisfactorily explained in terms of dust-reprocessing by Clavel,
Wamsteker \& Glass (1989), using a model put forward by Barvainis (1987). The
delay continues to exist in data obtained since the completion of the
earlier work.

In many cases, the variability is most apparent at the longer wavelengths,
especially $L$. This can have two causes, swamping at short wavelengths by
the underlying galaxy in the 12 arcsec diameter aperture (as mentioned) and
circumnuclear reddening, which is almost certainly present in NGC\,1068.

Some constraint on the size of the emitting regions can be obtained from the
timescale on which substantial variations can occur. If a dust re-emission
model is accepted, the material may be distributed at different
distances from the nucleus of the active galaxy and at arbitrary angles.
Thus, the infrared response to an instantaneous change in the ultraviolet
[e.g., $F_{\nu \rm{uv}}(t) \propto \delta(t)$] will usually be smeared out.
Only in the case of a ring of dust perpendicular to the line of sight can a
very sharp response be expected (Barvainis, 1992).

\section{Search for delays}

A number of the galaxies in the present paper have been monitored at shorter
wavelengths for various lengths of time and some of this work is included in
Fig 2. The variations in $U$ tend to have higher amplitude and possibly
shorter time scales than those at infrared wavelengths.

Delays between the short-wavelength light curve(s), thought to be indicative
of the UV flux from the central engine, and the $L$ flux originating in the
dust that it is heating, have been looked for using a simple
cross-correlation programme based on the interpolation method of Gaskell
\& Peterson (1987). It should be emphasized that there are many opinions as
to the best method for finding delays in cases where the data are
irregularly spaced. A more detailed analysis will be carried out in a future
paper.

\setcounter{figure}{0}
\begin{figure}
\epsfxsize=8.2cm
\epsffile[40 110 573 681]{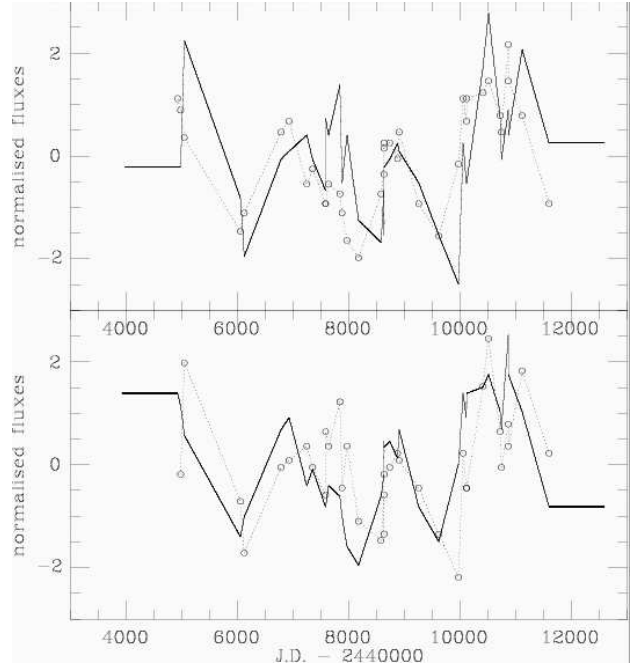}

\caption{Data for H0557-383 adjusted to have mean value 0 and rms
deviation 1, before cross-correlation. In the lower panel, the $J$ data are
interpolated and extrapolated and in the upper the $L$ data are so treated.
The interpolated data are given as solid lines and the non-interpolated data
are joined with dotted lines.}

\end{figure}

The Gaskell \& Peterson method was applied to those galaxies with
significant variability and many data points. Ideally, the data should have
been sampled at regular intervals, shorter than the expected delay times,
for this procedure to be reliable. The programme first interpolates the
longest wavelength photometry to give values for each day as well as
extrapolating its starting and stopping values as constants for 1000 days
before and after the series. Each light curve is shifted by a constant to
have average value 0 and divided so as to have rms deviation 1. The
cross-correlation is performed for the dates of the shorter-wavelength
observations, using a range of 1000 days in each direction. The correlogram
is divided by its maximum value so that the maximum of the peak is 1.0. 
Some of the causes that may distort the correlogram are (a) insufficient
sampling (b) long gaps in the data, particularly if these coincide with
enhanced activity (deviation from flat behaviour) (c) closely-clustered
measurements, which cause over-weighting of certain sections of the
non-interpolated sequences (d) an amplitude of variation that is not much
greater than the noise, such as occurs at the $J$ wavelength in some
galaxies (e) noise (f) correlated errors from sampling of the two
wavelengths quasi-simultaneously and (f) extrapolation of the data before
and after the series, as described. No attempt was made to correct the data
for these effects, except that some early and very sparsely sampled data
were removed in a few cases. Allowances for these problems will be made in
future analysis work, when possible. 

The procedure is carried out twice, exchanging the two data sets, so that in
the second case the shorter-wavelength data are interpolated instead of the
longer. The position of the peak is taken to be the centre of the
correlogram measured at the vertical level of 0.8 to avoid sometimes spiky
peaks. The results are presented in Table 2. The technique used is known to
be lacking in mathematical rigour, and can only be justified by the fact
that almost all galaxies tried show some degree of delay between the $U$ or
$J$ and the $L$ light curves. To check the input data the shifted and
divided data sets were plotted for inspection. In some cases the delay is
obvious to the eye; fig 1 is an example of such a plot, for H0557-383. The
$L$ vs $J$ cross-correlation gives 310d and the $J$ vs $L$ gives --300d.
Averaging these two implies a delay of about 305d. The longer delays are
likely to be the most certain and the shorter ones should be regarded as
indicative only. If the cross-correlation has multiple peaks at the 0.8
level or greater it is rejected. In the case of MCG-2-58-22 the
$J$ autocorrelation function was very broad, as were the cross-correlation
peaks, although suggestive of a delay of several hundred days.

There is no guarantee that the delay between the ultraviolet and the
infrared response is a constant quantity. The evaporation of dusty clouds
near an AGN has been considered by Pier \& Voit (1995) and it is clear
that a prolonged period of high activity will cause the inner edge of the
dust torus to retreat, increasing the propagation time for the UV radiation.
According to Barvainis (1992), for an equilibrium situation, the radius
within which particles will evaporate is

\[
R_{\rm evap} = 1.1
L^{0.5}_{46}e^{-\tau_{UV/2}}(T/1500\rm{K})^{-2.8}({\it a}/0.05\mu{\rm m})^{-0.5}
{\rm pc,}\]

\noindent where $L_{46}$ is the UV luminosity in units of $10^{46} {\rm erg}
s^{-1}$, $\tau_{\rm UV}$ is the UV optical depth of the cloud, $T$ is
the temperature of the hottest grains and $a$ is the grain radius in $\mu$m.

The present data show that long-term secular variations in the average
infrared luminosity of Seyfert galaxies are quite common. Their UV
luminosity is probably also variable, so that the Barvainis (1992) relation
given above should be applied with caution if at all.

A possibility for further work, pointed out by Dr V.L. Oknyanskij (private
communication), would be to concentrate on the $K$-band emission, which is
likely to be more representative of the hottest dust component and may show
a more precise response to ultraviolet variations than $L$.

\setcounter{figure}{1}
\begin{figure*}
\begin{minipage}{17.5cm}
\epsfxsize=17.5cm
\epsffile[119 141 494 650]{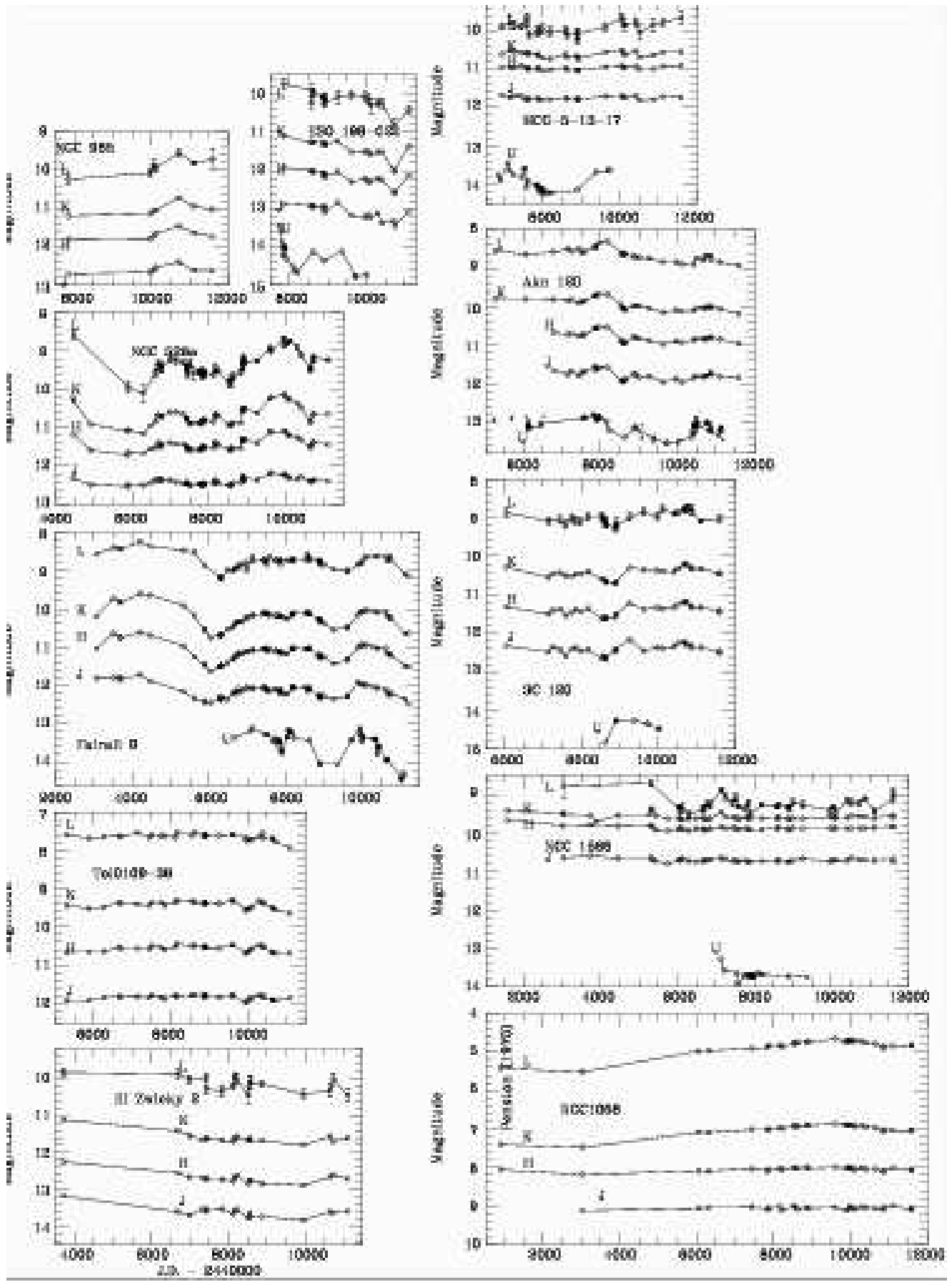}
\caption{(a) $JHKL$ light curves for Seyfert galaxies, plotted on uniform
magnitude and time scales.}
\end{minipage}
\end{figure*}

\setcounter{figure}{1}
\begin{figure*}
\begin{minipage}{17.5cm}
\epsfxsize=16.5cm
\epsffile[118 173 495 618]{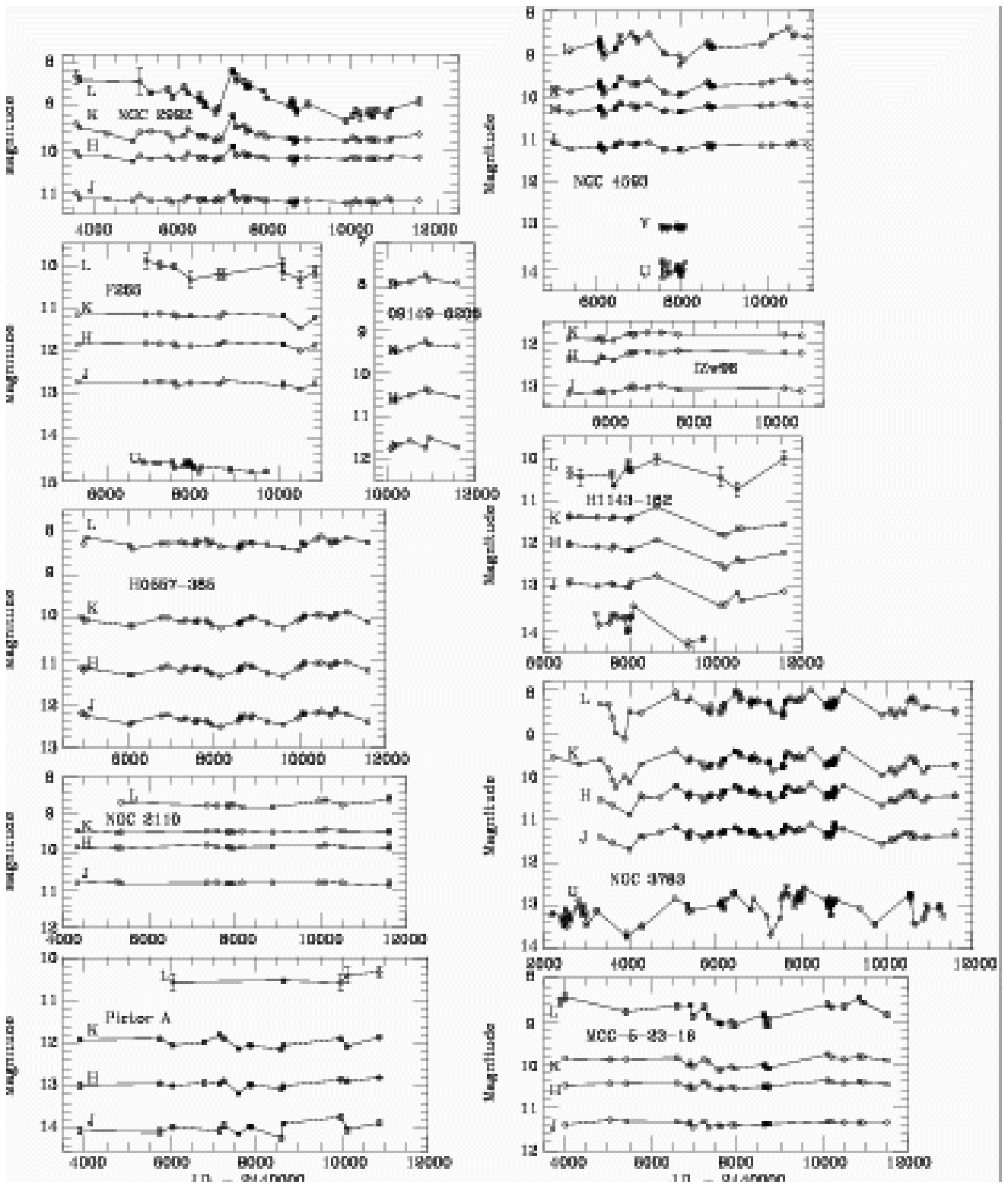}
\caption{(b) Light curves for Seyfert galaxies}
\end{minipage}
\end{figure*}

\setcounter{figure}{1}
\begin{figure*}
\begin{minipage}{17.5cm}
\epsfxsize=15.8cm
\epsffile[118 121 495 670]{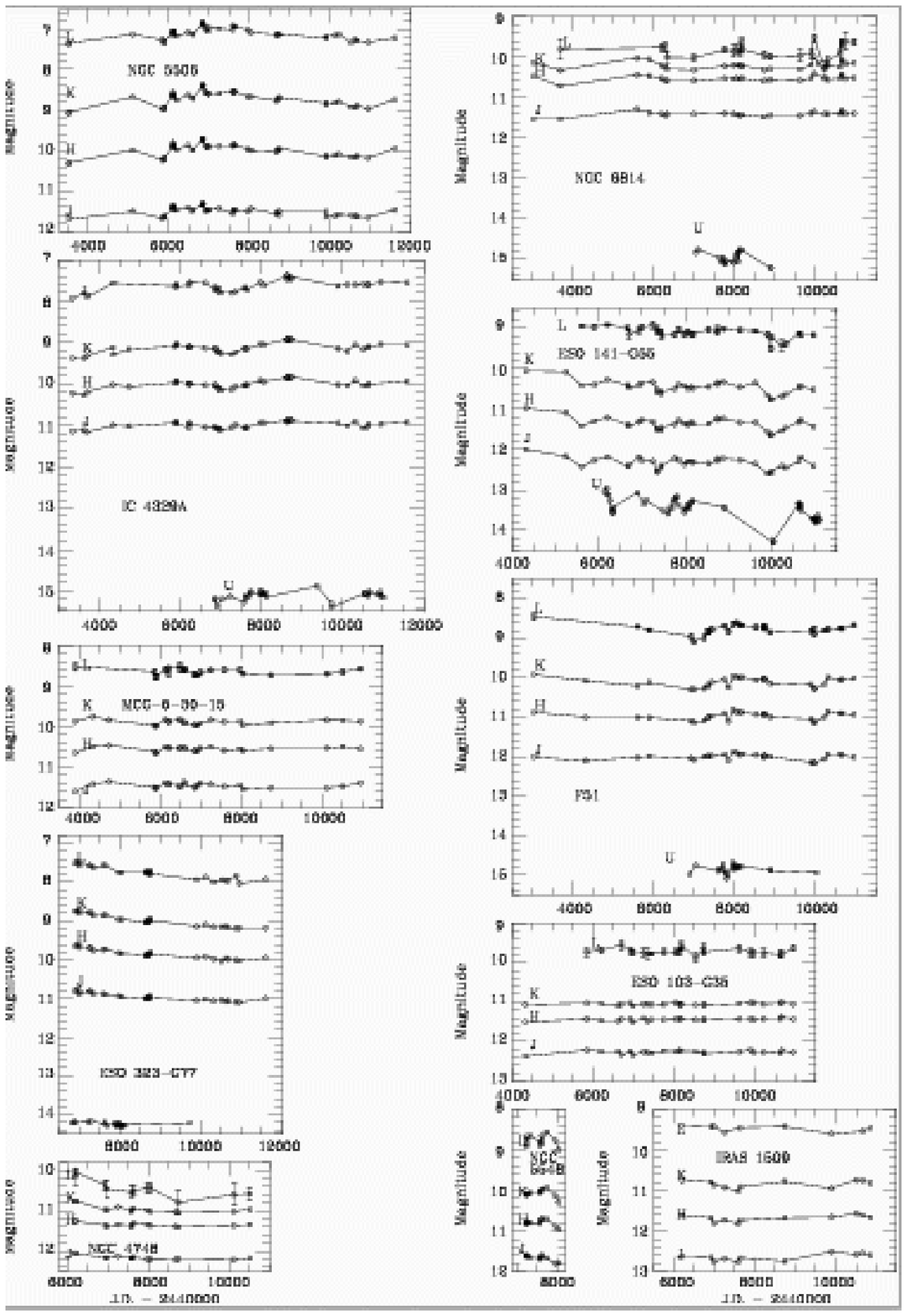}
\caption{(c) Light curves for Seyfert galaxies}
\end{minipage}
\end{figure*}

\setcounter{figure}{1}
\begin{figure*}
\begin{minipage}{17.5cm}
\epsfxsize=15.1cm
\epsffile[118 203 495 588]{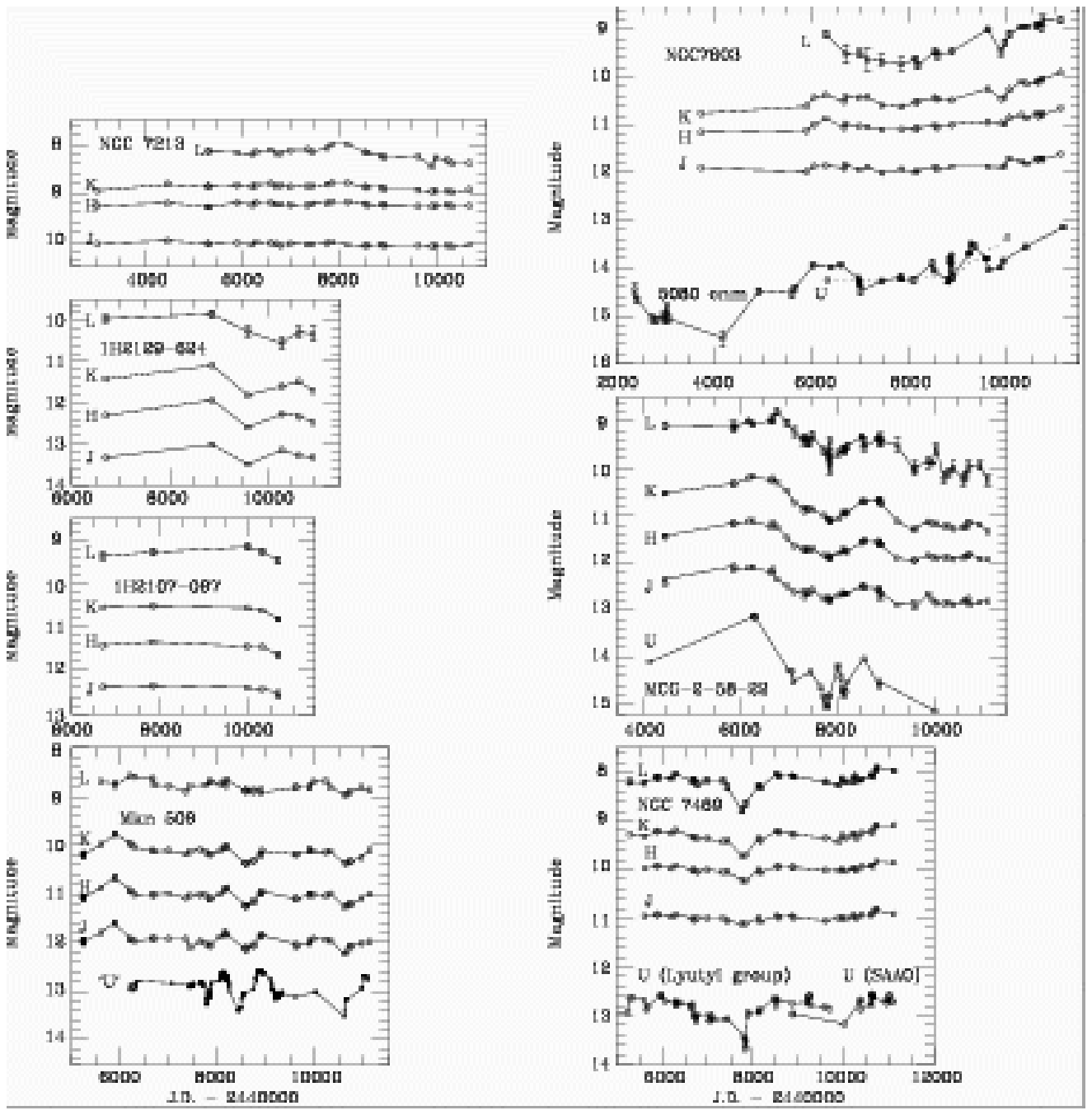}
\caption{(d) Light curves for Seyfert galaxies}
\end{minipage}
\end{figure*}

\begin{table}
\caption{Delays between $U$ or $J$ and $L$, determined by the cross-correlation
programme}
\begin{tabular}{llllll}
Name & bands & delay  & bands & delay  & average$^1$  \\
     &       & (days) &       & (days) & (days)       \\
III Zwicky 2   & -\\
Tol 0109-38    & $L,J$ & +100 & $J,L$ & -90  & 95  \\
F9             & $L,U$ & +500 & $U,L$ & -440 & 470 \\
F9             & $L,J$ & +370 & $J,L$ & -370 & 370 \\
NGC526A        & $L,J$ & +120 & $J,L$ & -250 & 185 \\
NGC985         & -                \\
ESO 198-G24    & -                \\
NGC1068        & -                \\
NGC1566        & -                \\
3C120          & $L,J$ & +160 & $J,L$ & -160 & 160 \\
Akn 120        & $L,J$ & +120 & $J,L$ &    0 &  60 \\
Akn 120        & $L,U$ & +250 & $U,L$ & -380 & 315 \\
MCG-5-13-17    & -                \\
Pic A          & -                \\
NGC2110        & -                \\
H0557-383      & $L,J$ & +310 & $J,L$ &-300 & 305  \\
F265           & -                \\
IRAS09149-6206 & -                \\
NGC2992        & $L,J$ &  0   & $J,L$ & -70 &  35  \\
MCG-5-23-16    & -               \\
NGC3783        & $L,J$ & 185  & $J,L$ & -110 & 148 \\
NGC3783        & $L,U$ & 200  & $U,L$ & -170 & 185 \\
H1143-182      & -                \\
I Zwicky 96    & -                \\
NGC4593        & $L,J$ & 110  & $J,L$ &   0  &  55  \\
NGC4748        & -                \\
ESO 323-G77    & -                \\
MCG-6-30-15    & $L,J$ & 47   & $J,L$ & -20  &  34  \\
IC4329A        & $L,J$$^2$ & 135   & $J,L$ & 70  &  33  \\
IC4329A        & $L,J$$^2$ & 150   & $J,L$ & -20  & 85   \\
NGC5506        & $L,J$ & -10  & $J,L$ & -50$^3$  &  25  \\
NGC5548        & -                \\
IRAS15091-2107 & -                \\
ESO 103-G35    & -                \\
F51            & $L,J$ & 20   & $J,L$ & -10  &  15  \\
ESO 141-G55    & $L,U$ & 180  & $U,L$ & -280 &  230 \\
ESO 141-G55    & $L,J$ & 210  & $J,L$ & -100 &  155 \\
NGC6814        & $L,J$ &  80  & $J,L$ &    0 &  40  \\
Mkn 509        & -                \\                
1H2107-097     & -                \\
1H2129-624     & -                \\
NGC7213        & -                \\
NGC7469        & $L,U$ & 30   & $U,L$ &  -12 &   21 \\
NGC7469        & $L,J$ & 110  & $J,L$ &  +30 &   40 \\
MCG-2-58-22    & $L,J$ & 270  & $J,L$ & -500 &  155 \\
NGC7603        & -                \\
\end{tabular}                    
\\
Notes: 

$^1$ The `average' column gives the average of the two `delay' columns, for
a delay in the sense long wavelength - short wavelength.

$^2$ The CCFs are asymmetric. At the usual level of 0.8 the delay is 33 d
and at 0.7 it is 85 d.

$^3$ The $J$ vs $L$ CCF is asymmetric. The delay is $\sim$ 0 at the peak.

\end{table}                      

\section{Spectral Energy Distributions of the Variable Components}

In the $UBVRI$ spectral region Winkler et al (1992) (see also Winkler, 1997)
found that the fluxes from many Seyferts, when plotted against each other on
linear scales, follow clear linear relationships. This effect had been
observed previously by Cho{\l}oniewski (1981) in NGC\,4151 and a number of
other active galaxies, QSOs and BL Lac objects. The implication is that the
spectral energy distribution of the variable component is constant with time
in a given galaxy and does not change with the level of activity. The slopes
obtained from the linear flux-flux diagrams can be expressed as colours and
it is found that they have some tendency to have values that are independent
of the galaxy being considered. Galaxies with nuclei that are clearly
perceived to be obscured have redder variable components.

The colours of the variable components have also been found to stay constant
or very nearly constant in the near-infrared (Glass, 1992), and a number of
examples have already been published: NGC\,3783 (Glass 1992), NGC\,1068
(Glass 1995), NGC\,2992 (Glass 1997a) and NGC\,7469 (Glass 1998). Figs
4a--f, 5 and 6 give data in the form of flux-flux plots for many of the
galaxies in this paper. The linear relations between most pairs of colours
are obvious. Only in the cases of III Zw 2, NGC2992, MCG-2-58-22 and NGC7603
do there seem to be changes of slope with intensity, in the $L$ vs $K$
diagram. These may be associated with long-term monotonic changes and are
discussed in more detail below.

The infrared colours of the variable components of all the galaxies in this
sample have been determined from the flux-flux plots. Each galaxy was first
dereddened according to the $E_{B-V}$ values given in Table 1 and the
magnitudes were converted to fluxes using the calibration for a zeroth mag
star of log(flux) = --22.82, --23.01, --23.21 and --23.55 Wm$^{-2}$Hz$^{-1}$
for $J$, $H$, $K$ and $L$ respectively. The procedure of Simon and Drake
(1989), which assumes that all the scatter is due to observational error,
was then used to calculate the regression coefficients $F_{HJ}$, $F_{KH}$
and $F_{LK}$ and their standard errors. The programme assumes constant
$JHK$ errors of 0.03 mag and $L$ errors of 0.06 mag, which were exceeded in
some cases.

The colours of the variable components were calculated using the following
formulae:

\[
J-H=2.5({\rm log}F_{HJ} +0.19)
\]  

\[
H-K=2.5({\rm log}F_{KH} +0.20)
\]  

\[
K-L=2.5({\rm log}F_{LK} +0.34).
\]

The colours and errors are given in Table 3. The errors can sometimes be
quite large. They arise from (a) the variations in many cases being quite
small (b) lower-quality photometry in the $L$-band, where the objects were
often rather faint and (c) from uncertainties in the telluric extinction
corrections at $L$ and to a lesser extent at $J$. (The $H$ and $K$
extinctions of the earth's atmosphere track each other very well, so that
the $H-K$ colour is the most reliable; Glass \& Carter, 1989).

\begin{table}
\caption{Slopes of the variable components of Seyfert galaxies, expressed as
colours.}
\begin{tabular}{llll}
Name & $H/J$ & $K/H$ & $L/K$\\
III Zw 2    &   0.96 $\pm$ 0.11  & 1.19 $\pm$ 0.11  & 1.62 $\pm$ 0.21 \\
Tol 0109-38 &   1.76 $\pm$ 0.32  & 1.35 $\pm$ 0.09  & 1.67 $\pm$ 0.30 \\
F9          &   1.29 $\pm$ 0.05  & 1.08 $\pm$ 0.04  & 1.20 $\pm$ 0.06 \\
NGC526a     &   1.68 $\pm$ 0.10  & 1.31 $\pm$ 0.04  & 1.67 $\pm$ 0.05 \\
NGC985      &   1.08 $\pm$ 0.24  & 1.03 $\pm$ 0.17  & 1.44 $\pm$ 0.19 \\
ESO 198-G24 &   0.91 $\pm$ 0.09  & 1.01 $\pm$ 0.07  & 1.35 $\pm$ 0.14 \\
NGC1068     &   1.79 $\pm$ 0.72  & 2.47 $\pm$ 0.36  & 2.48 $\pm$ 0.21 \\
NGC1566     &   1.19 $\pm$ 0.28  & 0.72 $\pm$ 0.16  & 1.83 $\pm$ 0.11 \\
3C120       &   1.06 $\pm$ 0.10  & 1.01 $\pm$ 0.09  & 1.38 $\pm$ 0.14 \\
Akn 120     &   1.14 $\pm$ 0.10  & 0.96 $\pm$ 0.08  & 1.33 $\pm$ 0.11 \\
MCG-5-13-17 &   0.70 $\pm$ 0.41  & 1.04 $\pm$ 0.23  & 1.63 $\pm$ 0.17 \\
Pic A       &   0.42 $\pm$ 0.34  & 1.63 $\pm$ 0.33  & 5.85 $\pm$ 0.54 \\
NGC2110     &   5.47 $\pm$ 0.54  & 0.14 $\pm$ 0.68  & 5.85 $\pm$ 0.54 \\
H0557-383   &   1.18 $\pm$ 0.16  & 1.11 $\pm$ 0.15  & 1.68 $\pm$ 0.40 \\
F265        &   0.66 $\pm$ 0.49  & 1.16 $\pm$ 0.44  & 5.85 $\pm$ 0.54 \\
IRAS09149-6206& 2.28 $\pm$ 1.05  & 1.00 $\pm$ 0.30  & 1.22 $\pm$ 0.54 \\
NGC2992     &   1.25 $\pm$ 0.12  & 1.27 $\pm$ 0.06  & 1.65 $\pm$ 0.05 \\
MCG-5-23-16 &   1.28 $\pm$ 0.34  & 1.32 $\pm$ 0.16  & 1.97 $\pm$ 0.13 \\
NGC3783     &   0.96 $\pm$ 0.07  & 1.04 $\pm$ 0.05  & 1.35 $\pm$ 0.08 \\
H1143-183   &   0.75 $\pm$ 0.09  & 0.70 $\pm$ 0.09  & 1.22 $\pm$ 0.32 \\
I Zw 96     &   1.32 $\pm$ 0.33  & 0.08 $\pm$ 0.22  & 5.85 $\pm$ 0.54 \\
NGC4593     &   1.31 $\pm$ 0.21  & 0.98 $\pm$ 0.10  & 1.55 $\pm$ 0.10 \\
NGC4748     &   0.69 $\pm$ 0.37  & 1.19 $\pm$ 0.23  & 1.63 $\pm$ 0.14 \\
ESO 323-G77 &   1.40 $\pm$ 0.14  & 1.07 $\pm$ 0.08  & 1.28 $\pm$ 0.10 \\
MCG-6-30-15 &   0.87 $\pm$ 0.32  & 0.92 $\pm$ 0.26  & 1.68 $\pm$ 0.39 \\
IC4329A     &   1.33 $\pm$ 0.14  & 1.04 $\pm$ 0.09  & 1.51 $\pm$ 0.14 \\
NGC5506     &   2.00 $\pm$ 0.16  & 1.42 $\pm$ 0.08  & 1.35 $\pm$ 0.10 \\
NGC5548     &   1.08 $\pm$ 0.40  & 0.94 $\pm$ 0.23  & 1.79 $\pm$ 0.39 \\
IRAS15091-2107& 0.94 $\pm$ 0.32  & 1.13 $\pm$ 0.30  & 1.05 $\pm$ 0.80 \\
ESO 103-G35 &   0.62 $\pm$ 0.70  &-0.09 $\pm$ 0.58  & 5.85 $\pm$ 0.54 \\
F51         &   1.37 $\pm$ 0.20  & 1.11 $\pm$ 0.13  & 1.39 $\pm$ 0.17 \\
ESO 141-G55 &   1.06 $\pm$ 0.09  & 1.02 $\pm$ 0.07  & 1.39 $\pm$ 0.17 \\
NGC6814     &   1.20 $\pm$ 0.35  & 0.87 $\pm$ 0.15  & 1.07 $\pm$ 0.12 \\
Mkn 509     &   0.96 $\pm$ 0.08  & 0.97 $\pm$ 0.07  & 0.49 $\pm$ 0.19 \\
1H2107-097  &   1.02 $\pm$ 0.25  & 1.04 $\pm$ 0.19  & 1.06 $\pm$ 0.39 \\
1H2129-624  &   1.35 $\pm$ 0.17  & 0.96 $\pm$ 0.11  & 1.22 $\pm$ 0.19 \\
NGC7213     &   1.10 $\pm$ 0.70  & 1.17 $\pm$ 0.32  & 1.48 $\pm$ 0.20 \\
NGC7469     &   1.22 $\pm$ 0.17  & 1.15 $\pm$ 0.10  & 1.31 $\pm$ 0.11 \\
MCG-2-58-22 &   1.04 $\pm$ 0.04  & 1.13 $\pm$ 0.03  & 1.30 $\pm$ 0.04 \\
NGC7603     &   1.21 $\pm$ 0.12  & 1.16 $\pm$ 0.06  & 1.52 $\pm$ 0.07 \\

\end{tabular}

\end{table}

The colours from Table 3 have been plotted in a histogram (fig 3). Those
with errors in excess of 0.4 mag have been omitted.  

\setcounter{figure}{2}
\begin{figure} 
\epsfxsize=8.2cm 
\epsffile[118 218 495 573]{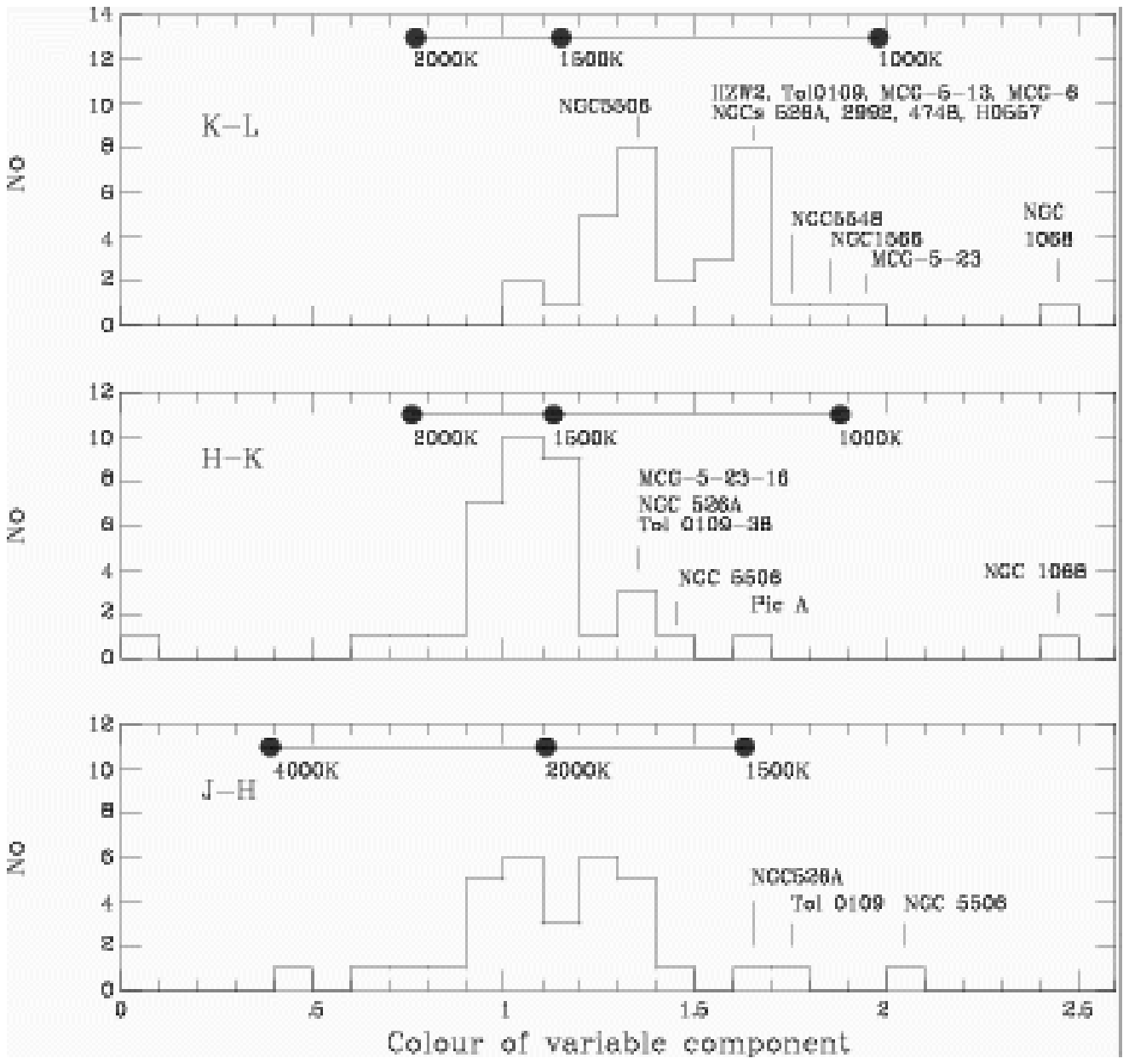}
\caption{Distribution of infrared colours of the variable components of
Seyfert galaxies. Blackbody colour temperatures are shown at the tops of
each panel. Measurements with errors $\geq$0.4 have been omitted.}
\end{figure}

There is a strong tendency for the colour temperatures to clump at
particular values. The best-determined colours are $H-K$, whose histogram
peak at 1.06 corresponds to a temperature around 1600K, believed to be the
highest temperature at which dust can exist without sublimation. These
wavelengths may be the least contaminated by the tail of the nuclear
ultraviolet continuum or an extra dust component. Taking only galaxies with
errors in $H-K$ $<$ 0.2, the average colour in the 0.95, 1.05 and 1.15
columns of the histogram is $H-K$ = 1.056 $\pm$ 0.016.

In $J-H$, the typical colour temperature is around 2000K. It is likely that
the tail of the ultraviolet component contributes to the $J$ band
(1.25$\mu$m) more than to the $H$ (1.65$\mu$m), leading to an apparently hotter
colour. This possibility is considered in detail for NGC\,3783, for which
$UBVRI$ data are available, by Glass (1992). In NGC\,3783 the contribution
to the $J$ band from the extrapolated $UBV$ flux is nearly equal to the flux
from the short-wavelength tail of the blackbody radiation from dust at
1500K.

In $K-L$, the colour temperature is lower, about 1300--1400K, and may
represent dust slightly further from the nucleus that that which
predominates in $H$ and $K$. It is possible that there is a series of zones,
proceeding outward from the nucleus, with decreasing dust temperatures. It
is interesting to note that an optically thick spherical dust shell having a
radius of 470d (as measured for Fairall 9) and a temperature of 1400K would
have an $L$-band luminosity of 3 10$^{24}$W Hz$^{-1}$. This is about 8 times
the observed luminosity given in table 1. Fairall 9 has $U-B$
$\sim$ --0.85, indicating that there is almost no circumnuclear absorption
along the line of sight. Thus part of the factor of 8 may result from a
dust distribution which is not spherical but possibly toroidal.

\subsection{Reddenings estimated from the variable components}

It is noticeable that the variable part of the flux from the Seyfert 2
galaxy NGC\,1068 is much redder in $H-K$ and $K-L$ than the average for
Seyferts as a whole. This is an indication that its nucleus is highly
obscured, with $E_{H-K}$ = 2.47 -- 1.056 = 1.414 $\pm$ 0.37, or $E_{B-V}$ =
7.6 $\pm$ 2.0 ($A_V$ = 23 $\pm$ 6) for a normal galactic extinction law(see
also Glass, 1997b). In the $J$ band its variation is minimal as almost no
nuclear $J$ flux can penetrate the dust barrier: the $J-H$ colour of the
variable part cannot be determined precisely enough to be useful.

Five other galaxies besides NGC1068 have $H-K$ colours of their variable
components which stand out from the average of 1.056 $\pm$ 0.016 in Fig
3. NGC5506 has E$_{H-K}$ = 1.42 $\pm$ 0.08 -- 1.056 $\pm$ 0.016 or
0.364 $\pm$ 0.08. For the galactic extinction law given above and taking
E$_{B-V}$ = 0.33 for A$_V$ = 1.0, NGC5506 would have E$_{B-V}$ =
2.0 $\pm$ 0.4. Similarly, for NGC526a, we have E$_{B-V}$ = 1.37 $\pm$ 0.22,
for MCG-5-23-16, E$_{B-V}$ = 1.4 $\pm$ 0.9 and for NGC2992 E$_{B-V}$ = 1.2
$\pm$ 0.3. These values, though not very precise, are generally higher than
Maiolino et al (2001) report for the same galaxies by applying the galactic
extinction law to the broad H lines and may lend some support to their
conclusion that the extinction law (which they discuss in terms of
$E_{B-V}/N_{\rm H}$) in the vicinity of active galactic nuclei
is anomalous, i.e. that $A_V/E_{B-V} > 3.03$, the standard galactic value.
The fifth galaxy, not included in Maiolino et al (2001), is Tol 0109-36 with
E$_{B-V}$ = 1.6 $\pm$ 0.5.

\section{Constant flux component}

It is reasonable to assume that the flux in the 12 arcsec diameter aperture
has components other than the variable one, presumed here to be from
radiation by hot dust. In the simplest model it is required that this
remnant flux should be `ordinary' or inactive galaxy, i.e., a typical
mixture of late-type giant stars (as seen in the near-infrared). The $J-H$
and $H-K$ colours of such galaxies are remarkably independent of spectral
type (Glass, 1984). The effect of relativistic K-corrections have been
estimated from $K_{J-H} = 0.5z$, K$_{H-K} = 3.0z$, K$_{K-L} = --1.4z$ (see
Frogel et al, 1978; Longmore \& Sharples, 1982; Griersmith et al, 1982) and
used to predict the colours of such a component in any of the present
sample.

Returning to the flux-flux plots, Figs 4a--f, 6 and 7, if nuclear activity
were to cease altogether, the lowest point the fluxes could reach would be
given by the intersection of the observational line with the locus of
inactive galaxy flux. It is thus possible to separate the variable and fixed
components independently of model surface brightness profiles.

\subsection{Underlying galaxy component}

If there were no other components besides ordinary galaxy and the variable
source present in the 12-arcsec aperture, it would be expected that the
fluxes derived from the intersections in each panel should agree with each
other; i.e., that the $H$ flux of underlying galaxy from the $J$ vs $H$
flux-flux plot should agree with the $H$ flux from the $H$ vs $K$ plot and
the $K$ flux from the $H$ vs $K$ should agree with the $K$ flux from the $K$
vs $L$ plot. However, the accuracy with which the points of intersection can
be determined is limited by how well the slope of the variable component is
known and by the extent to which the observed line must be extrapolated to
meet the underlying galaxy. Further, in the $J$ vs $H$ plot, the accuracy is
also affected by the fact that the regression line is usually almost
parallel to the underlying galaxy line.

Some galaxies have solutions for $H$ and $K$ that are reasonably consistent.
These objects are usually of low luminosity with variable fluxes that are
much smaller than the underlying galaxy components, though tending to be a
larger fraction at the longest wavelength. Some idea of the errors to be
expected in the determination of the points of intersection can be obtained
by plotting the regression lines for the observed points with slope
increased by $\pm1\sigma$ and hinged about the average flux levels.

The flux-flux diagrams also reveal that some of the most luminous members of
the sample, such as H0557-383, H1143-182 and Mkn509, have almost no normal
galaxy contribution to the fluxes in the 12 arcsec diameter aperture; i.e.,
they are essentially QSOs.

\subsection{Other constant components}

\begin{figure*} 
\begin{minipage}{17.5cm} 
\epsfxsize=17.5cm 
\epsffile[41 248 570 543]{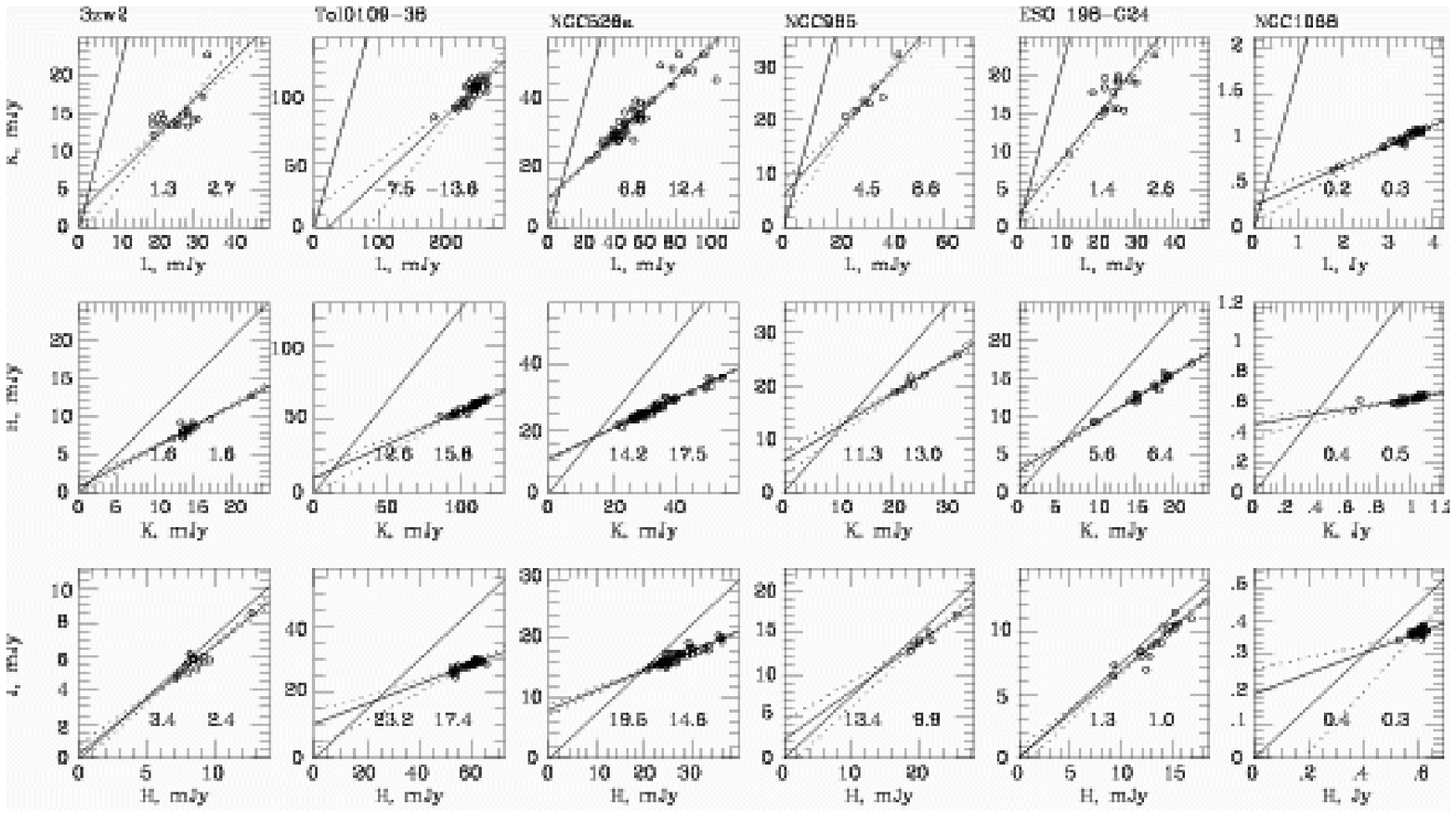} 

\caption{(a) Linear flux-flux diagrams of galaxies with variable nuclei.
The ranges of the abscissas and ordinates have been chosen to have the same
ratios from galaxy to galaxy in order to facilitate the comparison of slopes.
The regression lines are shown (solid) with companion lines having slopes
differing by $\pm\sigma$ (dotted). Also given for each diagram is a line
representing the colour of an average `ordinary galaxy' at the same
redshift. The coordinates of the intersection points are given. If the sole
non-variable content is ordinary galaxy, the $H$ coordinate of the
intersections should be the same for the $J$ vs $H$ and $H$ vs $K$ diagrams.
Similarly for the $K$ coordinates in the $H$ vs $K$ and $K$ vs $L$ diagrams.
The fact that this is not usually the case implies that a cool dust
component may also be present (see text). Note that error bars have not been
plotted, but errors may be judged from Fig 1.}

\end{minipage}
\end{figure*}

\setcounter{figure}{3}
\begin{figure*} 
\begin{minipage}{17.5cm} 
\epsfxsize=17.5cm 
\epsffile[41 248 570 543]{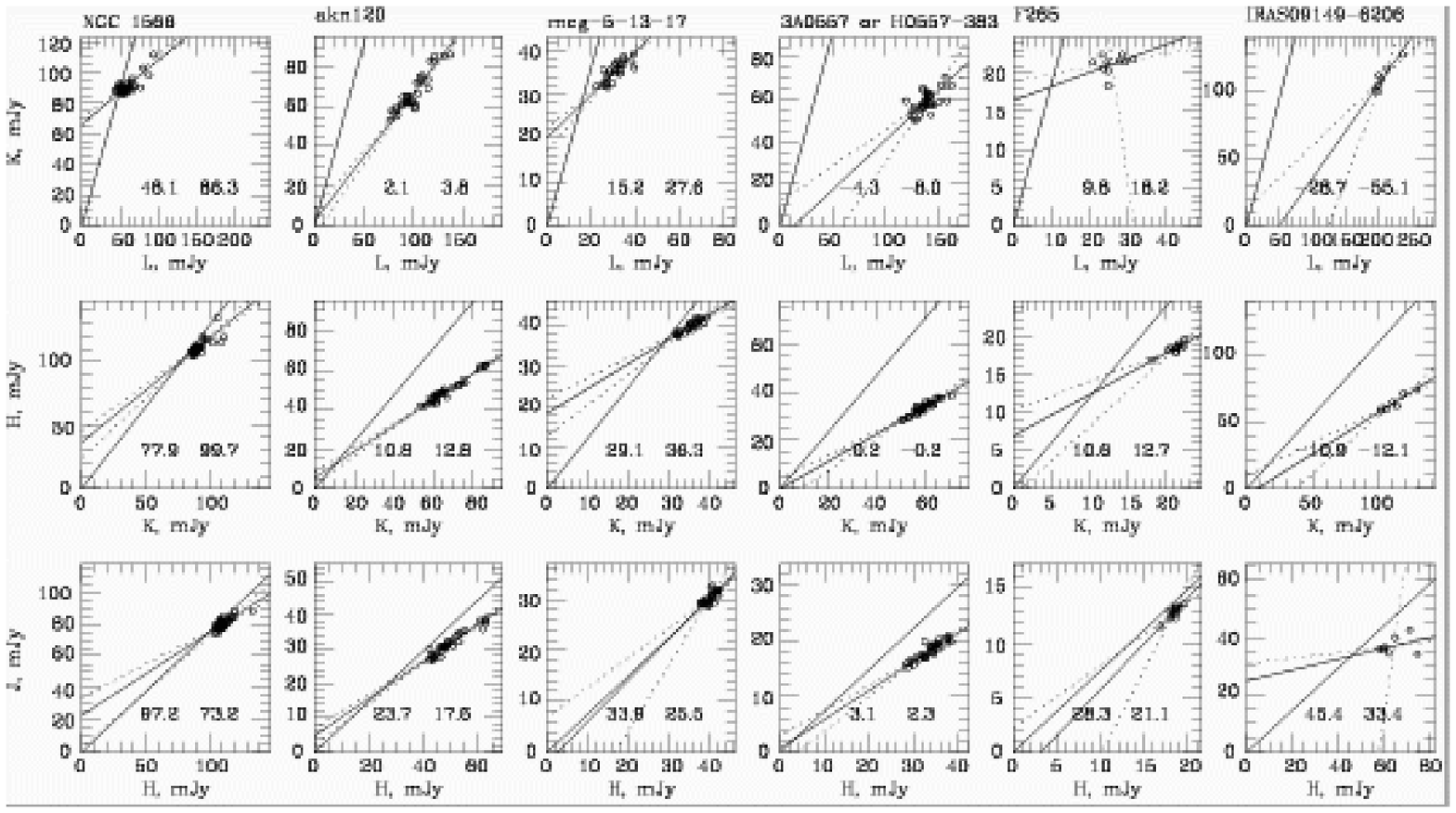} 
\caption{(b) Linear flux-flux diagrams (contd.)}
\end{minipage}
\end{figure*}

\setcounter{figure}{3}
\begin{figure*} 
\begin{minipage}{17.5cm} 
\epsfxsize=17.5cm 
\epsffile[41 248 570 543]{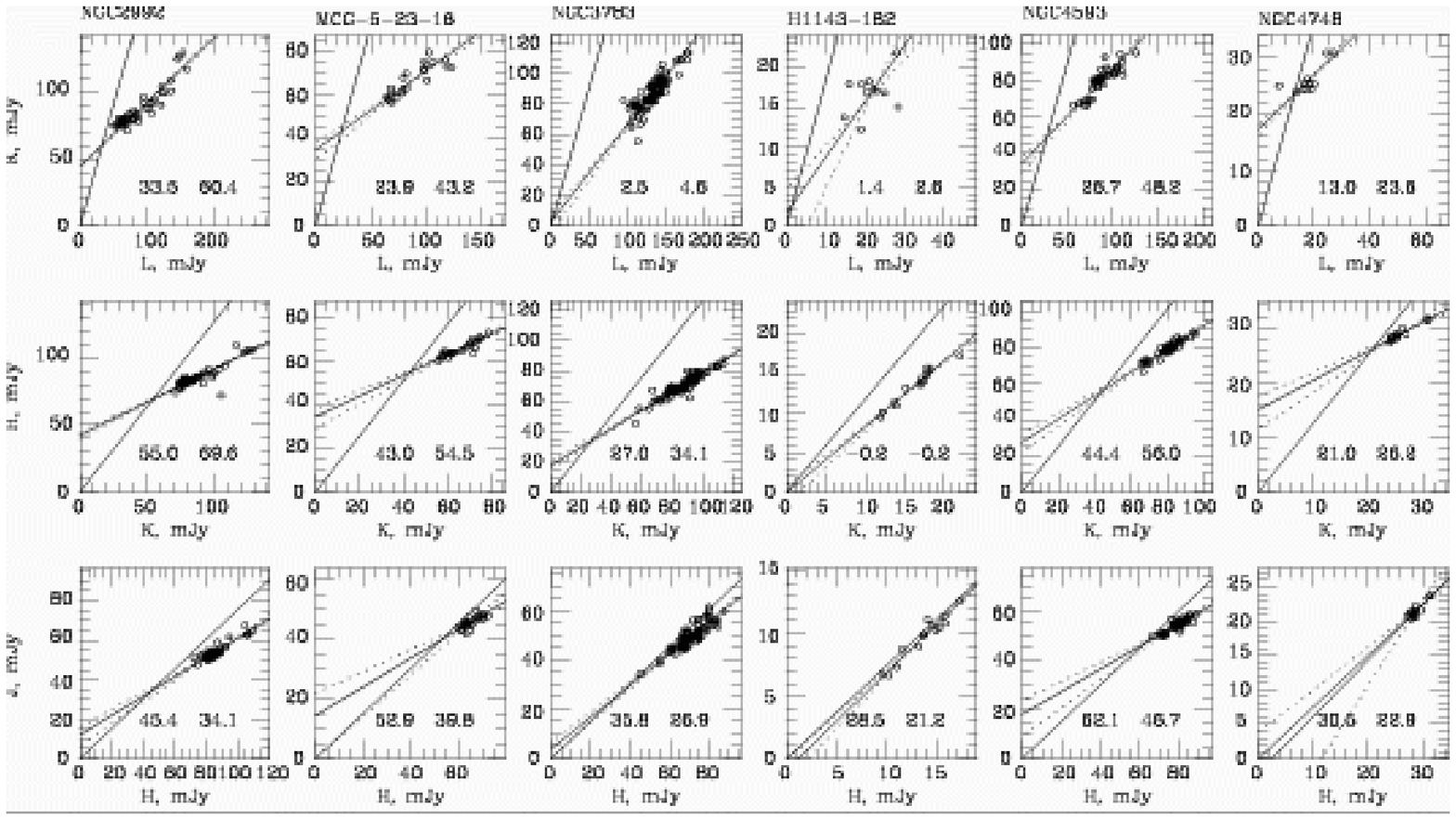} 
\caption{(c) Linear flux-flux diagrams (contd.)}
\end{minipage}
\end{figure*}

\setcounter{figure}{3}
\begin{figure*} 
\begin{minipage}{17.5cm} 
\epsfxsize=17.5cm 
\epsffile[41 248 570 543]{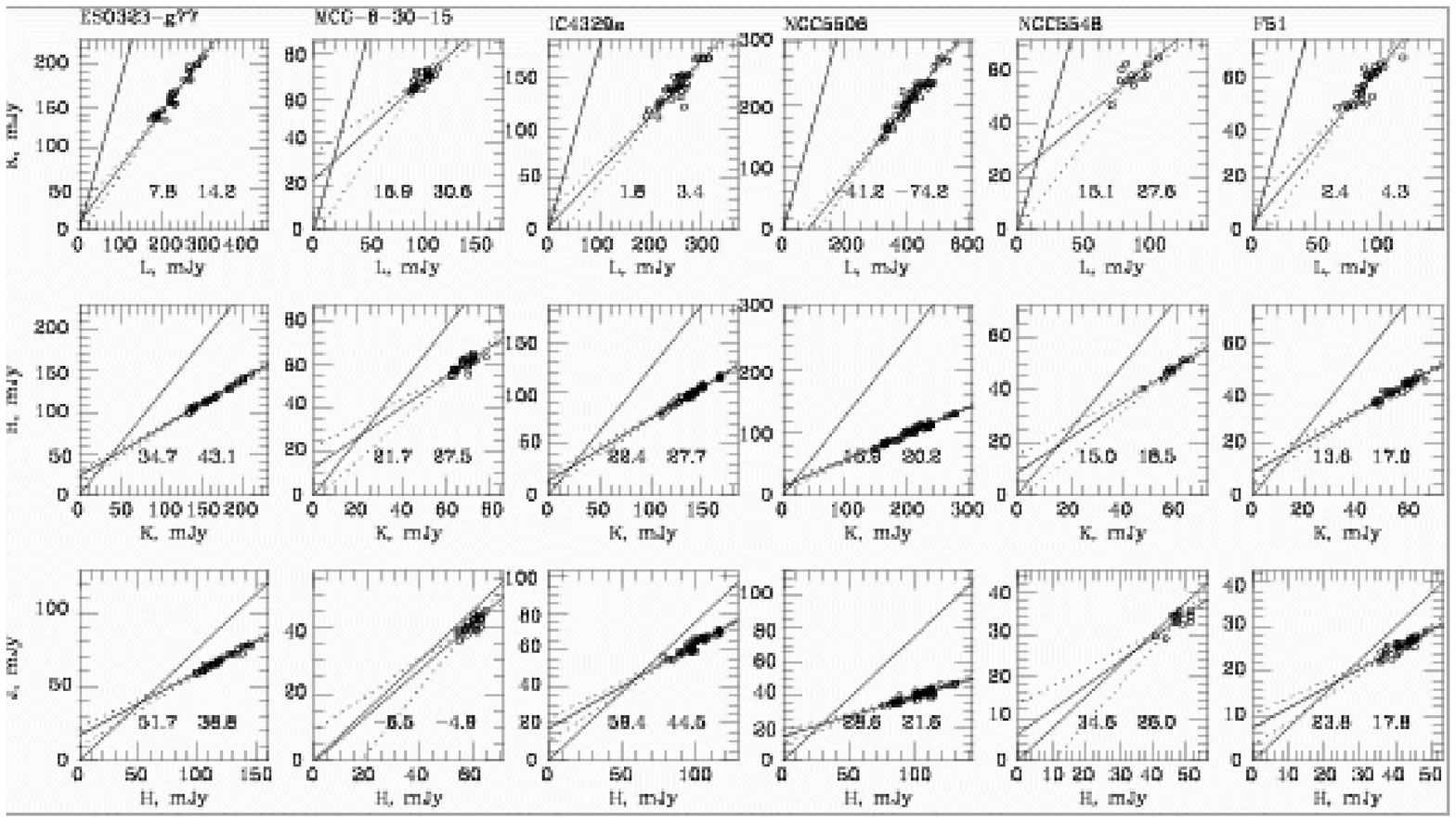} 
\caption{(d) Linear flux-flux diagrams (contd)}
\end{minipage}
\end{figure*}

\setcounter{figure}{3}
\begin{figure*} 
\begin{minipage}{17.5cm} 
\epsfxsize=17.5cm 
\epsffile[40 248 570 543]{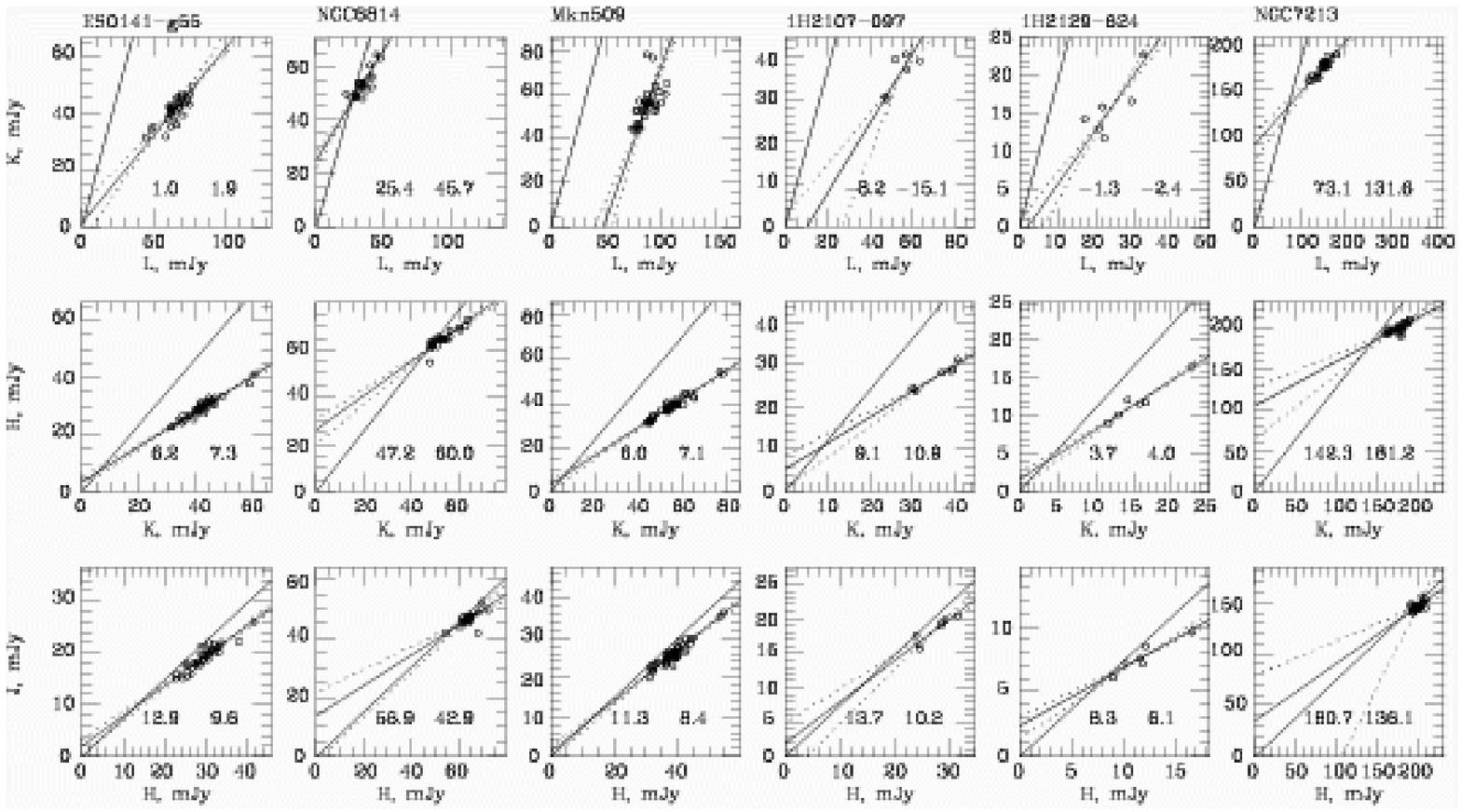} 
\caption{(e) Linear flux-flux diagrams (contd)}
\end{minipage}
\end{figure*}

Most of the Seyferts in the sample cannot be resolved into a variable
component and a simple underlying standard inactive galaxy.

Taking F9 (Fig 5) as an example, which will be discussed below in more
detail, the $H$ flux of the intersection in the $J$ vs $H$ diagram is 18 mJy,
whereas it is only 10.2 mJy in the $H$ vs $K$ diagram. Similarly, the $K$
flux of the intersection is 10.2 mJy in the $H$ vs $K$ and --21 mJy in the
$K$ vs $L$ diagrams, the latter being clearly unphysical. It will be seen
that these discrepancies can be resolved by postulating the existence of a
component with associated cool dust that contributes to the fixed flux,
increasing with wavelength.

Another example of this kind is NGC\,7469, which was discussed in detail by
Glass (1998). A circumnuclear ring exists in this galaxy and appears to
constitute a non-variable contribution in addition to the ordinary galaxy
stellar component. The ring has been measured at high spatial resolution in
$JHK$ by Genzel et al (1995), so that its contribution is known.

\section{Parameterization of the nuclear activity}

Because of the linear relations between fluxes at different wavelengths, the
position of a measured point along the regression line in a flux-flux
diagram (see fig 5) can be taken as a measure of the activity at the time.
We can define an activity parameter $A$ which rises from 0 at the point
where the regression line crosses the locus of the ordinary galaxy component
to 1 at the maximum of the observed flux. An error can be estimated for $A$
by determining the points at which the $\pm \sigma$ regression lines
intersect with the same locus. The error can be large if there are only a
few observational points or if the regression line is almost parallel to the
ordinary galaxy line. Such is often the case for the $H,J$ diagram.

For several galaxies, the errors are small and the values of $A$ found from
all three diagrams agree reasonably well. These are well-observed objects
whose constant component appears to be pure inactive galaxy.

In the case of those galaxies where an extra constant component seems to be
present, the intersections in the $L,K$ diagram are lower down than
expected, and sometimes at unphysical values, because the extra component is
cool and causes a displacement of the longer-wavelength fluxes (plotted
always along the x-axis) to the right. A self-consistent picture of the
different contributions can often be derived by starting with the $H$ vs
$J$ or the $K$ vs $H$ regressions, according to the probable errors, and
estimating the effects of the extra component, followed by iterating.
Because of its low temperature it has relatively little effect at shorter
wavelengths. Based on the well-observed galaxy F9, the ratio of $L:K:H$
fluxes are about $4:1:0$, corresponding to a cool dust component of
$\sim$850K. This galaxy is discussed in detail in section 9.3. Its
$J$ and $H$ fluxes lead to a value of $A$ at the lowest level of
activity of about 11 $\pm$ 3\% of the maximum observed values.

Table 4 gives the value of A based on the $H$ vs $K$ diagrams for suitable
candidates, together with the underlying galaxy flux in the 12 arcsec
measuring aperture at $K$ and an estimate of the excess cool component, if
present, at $L$.

\begin{table} 

\caption{Further results from flux-flux diagrams: Activity parameters
and underlying galaxy components (within 12 arcsec diameter aperture) for
galaxies with sufficient variability and sufficient data.}

\begin{tabular}{llll}
Galaxy      & A$_{min}^{HK}$       & $K_{gal}$ (mJy)  & comments \\
III Zw 2    & 0.49$^{+.1}_{-0.06}$ & $<$3             & VS\\
Tol 0109-38 & 0.70$^{+.06}_{-.09}$ & $<$20            & \\
F9          & 0.11$^{+.04}_{-.02}$ & 16$\pm$2         & \\
NGC526a     & 0.056$^{+.025}_{-.027}$ & 14$\pm$2      & no strong CE \\
ESO198-G24  & 0.45$^{+.1}_{-.1}$   & 5.6$\pm$1        & no strong CE \\
NGC1068     & 0.33$^{+.05}_{-.02}$ & 390$\pm$50       & CE = 0.25Jy \\
NGC1566     & 0.27$^{+.29}_{-.08}$ & 87$\pm$5         & \\
3C120       & 0.5                  & $\sim$10         & CE = 16 mJy \\ 
Akn120      & 0.58$^{+.09}_{-.04}$ & $\sim$11         & CE = 35mJy \\
MCG-5-13-17 & 0.26$^{+.18}_{-.14}$ & 29$\pm$5         & \\
H0557-383   & 0.72                 & $<$7             & \\
NGC2992     & 0.20$^{+.04}_{-.02}$ & 55$\pm$2         & VS\\
MCG-5-23-16 & 0.35$^{+.05}_{-.05}$ & 43$\pm$2         & \\ 
NGC3783     & 0.28$^{+.04}_{-.04}$ & 32$\pm$5         & CE = 32 mJy\\
PKS1143-182 & 0.55$^{+.04}_{-.06}$ & $<$2             & \\
NGC4593     & 0.41$^{+.07}_{-.03}$ & 46$\pm$4         & no strong CE \\
NGC4748     & 0.28$^{+.10}_{-.06}$ & 21$\pm$2         & no strong CE \\
ESO323-G77  & 0.56$^{+.04}_{-.10}$ & 36$\pm$10        & no strong CE \\
MCG-6-30-15 & 0.74$^{+.08}_{-.09}$ & $<$30            & \\
IC4329A     & 0.37$^{+.17}_{-.08}$ & $\sim$50         & CE = 97 mJy \\
NGC5506     & 0.43                 & $\sim$27         & CE = 140 mJy \\
F51         & 0.45$^{+.16}_{-.10}$ & $\sim$17         & CE = 23 mJy \\
ESO141-G55  & 0.32                 & $\sim$12         & CE = 20 mJy \\
NGC6814     & $\sim$ 0             & 47$\pm$2         & no strong CE \\
MKN 509     & 0.51                 & $\sim$23         & CE = 40 \\
NGC7213     & 0.74$^{+.08}_{-.09}$ & 140$\pm$10       & no strong CE \\
NGC7469     & 0.19                 & $\sim$58         & CE = 41\\ 
MCG-2-58-22 & 0.15                 & $\sim$11         & CE = 15; VS \\
NGC7603     & 0.13$^{+.03}_{-.03}$ & 26$\pm$1         & no strong CE, VS \\
\end{tabular}

Notes:

CE denotes cool excess component at $L$, in mJy.

VS denotes variable $K$ vs $L$ slope

Values without quoted errors are the (approximate)
results of iterative solutions.

Absorption within the Milky Way galaxy has been allowed for.

\end{table}
 
\section{Comments on individual galaxies}

Many of the galaxies observed during this programme are the subjects of
extensive studies only a few of which will be referred to.

\setcounter{figure}{3}
\begin{figure} 
\epsfxsize=8.2cm 
\epsffile[116 194 495 597]{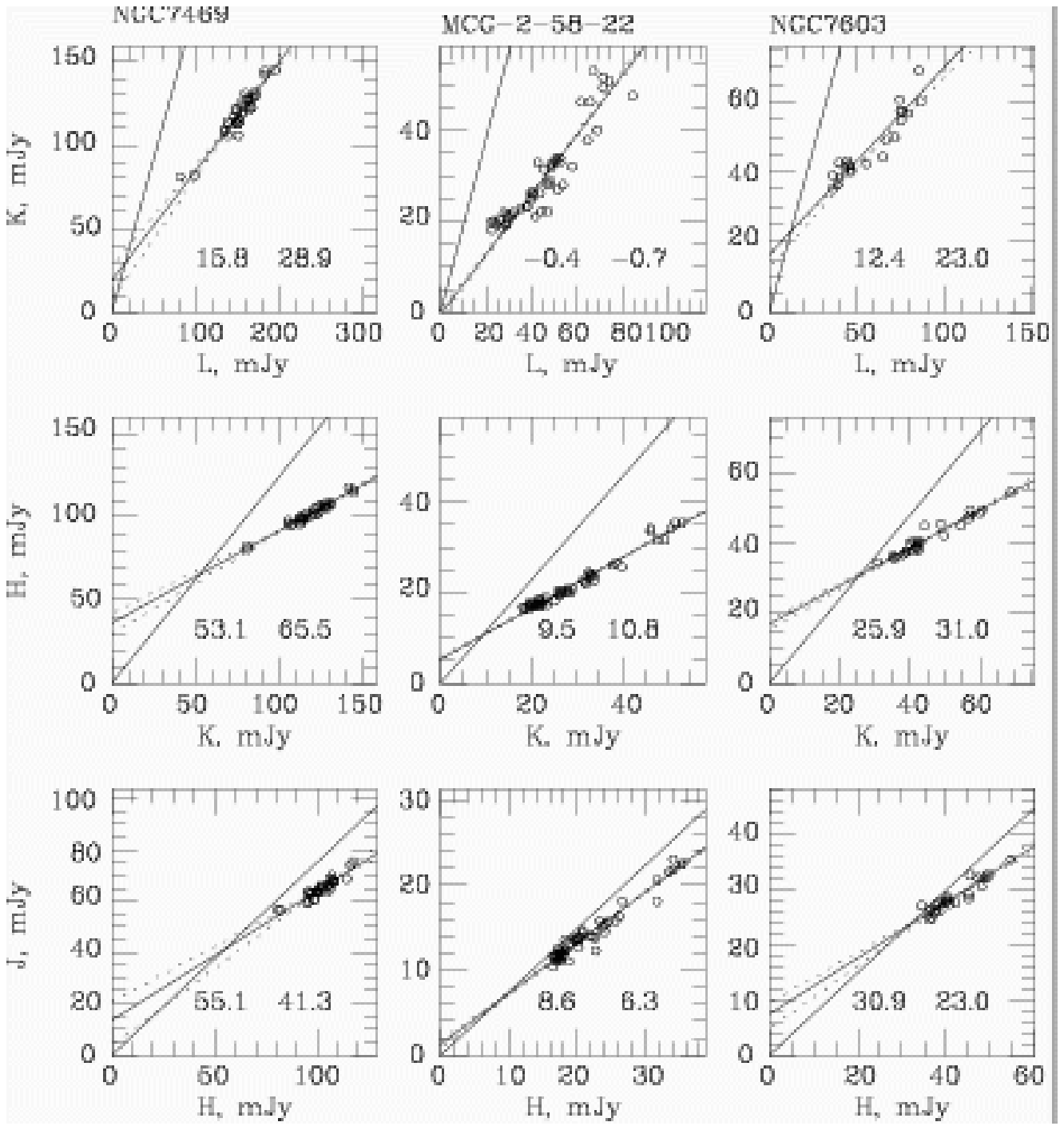} 
\caption{(f) Linear flux-flux diagrams (contd.)}
\end{figure}

\subsection{III Zw 2}

This is one of the galaxies in which variations were found by Lebofsky \&
Rieke (1980). At the time of their work (1978-79) it was at the high end of
its luminosity range. Their photometry falls (after correction for
reddening) close to the regression lines derived from the present work. The
infrared increase they observed was attributed to a previously noted
increase in its optical-UV radiation, providing early evidence for the dust
reverberation model.

The slope of the regression in the $K$ vs $L$ diagram may have changed with
time in the sense that the $L$ flux is continuing to decrease even though
the $K$ flux has bottomed out. This may be a reflection of a very long-term
decrease in the $L$ flux, arising from relatively distant dust, following a
protracted decline in UV activity since the time of the Lebofsky \& Rieke
work. See also NGC\,2992, MCG-2-58-22 and NGC\,7469 below.

\subsection{Tol 0109-28}

This Seyfert 2 galaxy has shown very little activity. The colours of its
variable components are considerably redder than average, suggesting strong
nuclear reddening. There appears to be a delay of about 100d between $J$ and
$L$.

\subsection{Fairall 9}

The data on Fairall 9 yield the best argument so far for the dust
reverberation model. The delay of $\sim$ 400 d found by Clavel, Wamsteker \&
Glass (1989) between the IUE continuum and the $L$-band light curve remains
the strongest evidence for the dust reverberation model.  Monitoring has
continued at SAAO in the $U$ band and a delay is still seen in the infrared
response. The SAAO $U$ data are shown in Fig 1. Variations seen in the new
$U$ data are found to precede those at $L$ by 470d. The lag between $J$ and
$L$ for the entire data set is found to be 370d.

Fig 5 shows the flux-flux plots for F9, together with the `activity
parameter', $A$. In the $J$ vs $H$ diagram, the point corresponding to the
level of activity the lowest flux comes closer to the `ordinary galaxy' line
than in the other two diagrams. We now assume that the point of intersection
of the two lines, $(H,J)$ = (18$\pm$2, 13.5$\pm$2), represents pure galaxy
and an activity of $A$ = 0. The minimum observed activity, as a proportion
of the maximum, is then $A = 0.11 \pm 0.03$.

Using the minimum and maximum levels of $A$, we can estimate the points on
the $H$ vs $K$ diagram where $A$ would be zero, namely at $(K,H)$ = (22,18),
with correlated uncertainties of about $\pm$2 mJy in each coordinate. It is
clear that in Fairall 9, as well as many other galaxies, this point does not
coincide with the intersection of the regression line and the ordinary
galaxy line. Instead, the position of the point at (22,18) can be regarded
as the sum of an `ordinary galaxy' component of (16,18) and an underlying
constant component with $F_H$ $\sim$ 0 (see next paragraph) and
$F_K$ = 6$\pm$2 mJy.

Similarly, for the $K$ vs $L$ diagram, the position of the point corresponding to
$A$ = 0, namely $(L,K)$ = (50,24) can be regarded as the sum of an ordinary
galaxy component of (9,16) with an underlying constant component of
(41,8). This corresponds to a $K-L$ colour of 2.42, or a blackbody
temperature of $\sim$850K. The $H-K$ colour for this temperature of
blackbody is 2.4, which would imply an $H$ flux only about 18\% of the $K$
flux, or about 1.4 mJy. This is sufficiently close to the estimated flux of
0 to be acceptable.     

Barvainis (1992) made a detailed reverberation model of Fairall 9, based on
the data presented by Glass (1986) and Clavel, Wamsteker \& Glass (1989). It
should be noted that information from a well-sampled IUE UV light curve,
contemporary with the infrared data, was available for this galaxy. From
Fig.\ 2 of Glass (1986) he took the underlying galaxy contribution in a 12
arcsec diameter aperture to be 14, 19, 17 and 9 mJy at $JHKL$ respectively.
These figures are close to the 13.5, 18, 16 and 9 mJy used in the work
just described. From fits of his dust model, Barvainis concluded that there
must be a `fourth component' of flux that is constant with time. He found
for this flux the $JHKL$ values 1.5, 3.7, 9.0 and 37mJy respectively,
arising from quiescent dust at about 800K or a power law with $\alpha$ = --3.
His deduction is consistent with the results found here, i.e. that the
fourth component has $F_K$ = 8 and $F_L$ = 34 mJy. The present method gives
no information as to $F_J$ and $F_H$ as they are tacitly taken to be zero
but, if the source is dust at 800K, we have seen that the fluxes at these
wavelengths will be small.

\begin{figure}
\epsfxsize=6cm
\epsffile[210 103 401 690]{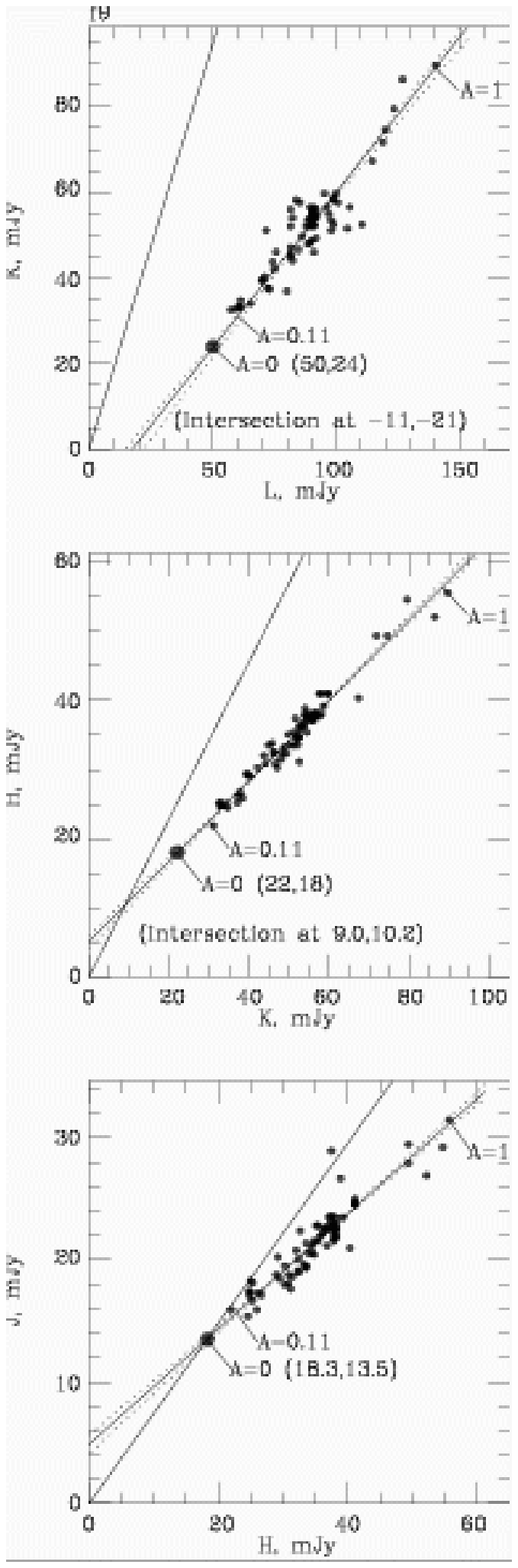}
\caption{Flux-flux diagrams for Fairall 9 observed through a 12-arcsec
diameter aperture, including information on the `Activity parameter'. $A$ is
taken to be 1 at the maximum observed level and to be zero at the place in
the $H$ vs $J$ diagram where the regression line intersects with the
ordinary galaxy line. The lowest observed fluxes in the same diagram have
$A=0.11 \pm 0.03$. The position of $A$=0 in the other two diagrams can be
estimated from the highest and lowest observed fluxes. From these the
contribution of an additional non-varying component of flux can be estimated
(see text).}
\end{figure}

\subsection{NGC\,526A}

This object appears heavily obscured in visible light and its variable
component is significantly redder than average. There is some evidence for a
delay between $J$ and $L$ of 200$\pm$100 days.

\subsection{ESO 198-G24}

A large visual range was found for this galaxy by Winkler et al (1992).

\subsection{NGC\,1068}

The archetype S2 galaxy, NGC\,1068 has been discussed in detail by Glass
(1995). The $J-H$ colour of its variable component is not well-determined,
but the $H-K$ and $K-L$ colours are by far the reddest of the sample.

\subsection{NGC\,1566}

A detailed discussion of the event around JD 2447000, as observed at many
wavelengths from the x-ray to the IR, was given by Baribaud et al (1992). A
time lag of 2 $\pm$ 1 months was found between the UV and the IR.  This
result was criticised by Oknyanskij \& Horne (2001), who find that the delay
is consistent with 0 and is less than 20d.

\subsection{3C\,120}

Lebofsky \& Rieke (1980) reported a decline in the $K$ flux of 3C\,120 from
levels of 64--71 mJy, that had been observed in 1970--1972 by Penston et al
(1974), to $\sim$24 mJy in 1978. This decline took at least three years.  In
the present work the (not dereddened) $K$ flux varied from 32 to 47 mJy.
Only three galaxies in the present sample have shown such large changes.
However, the case for 3C\,120 seems to be well corroborated and is almost
certainly real. The Penston et al and Lebofsky \& Rieke values are plotted
with the present data in Fig 6 as crosses. The fluxes tend to lie quite
close to the regression lines shown except at $L$, where the Penston $L$
values seem too bright. However, the standard errors of his measurements
were quite high.

\begin{figure}
\epsfxsize=8.2cm
\epsffile[39 155 574 636]{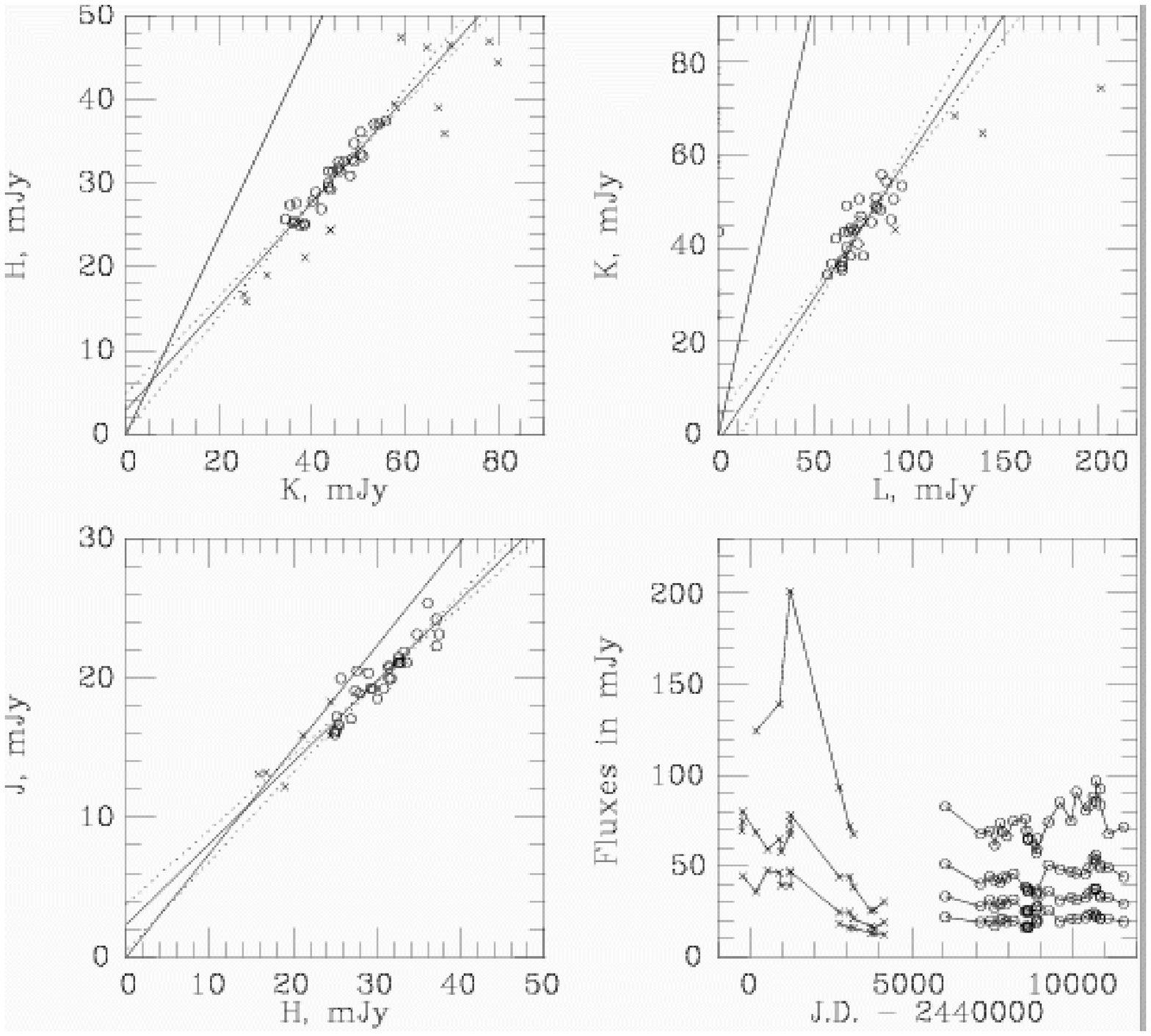}

\caption{Flux-flux diagrams and light curves for 3C120. The crosses are data
from other authors. Note that the errors on the early $L$ data range from 15
to 30 percent. The early data (shown by crosses) were not used in the
determination of the regression lines.}

\end{figure} 

Long-term $B$ light curves of 3C\,120 have been presented by Hagen-Thorn
et al (1997) and Clements et al (1995), in which a decline over the years
1973 to 1978 is clearly seen. 

A cross-correlation analysis between $J$ and $L$ shows that the longer
wavelength emission is delayed by about 160 days compared to the shorter.

\subsection{Akn 120}

Akn 120 has been monitored in the visible region at SAAO by Winkler et al
(1992), Winkler (1997) and in a service observing programme at SAAO under
the control of Dr D. Kilkenny.

It has also been observed extensively by Doroshenko \& Lyuty (1999), who
found that the $U-B$ colour of its variable component did not remain
constant during two episodes of fading. 

A delay of about 315 days was found between $U$ and $L$, but between $J$ and
$L$ only $\sim$60d was found. Clavel, Wamsteker \& Glass (1989) found that
the delay between $U$ and the near-IR wavelengths tended to be increase with
wavelength in F9. Uneven sampling could also play a part in the seeming
discrepancy.

\subsection{MCG-5-13-17}

Visible-region photometry from Winkler et al (1992) \& Winkler (1997) is
also shown. The $U$ data are insufficient for cross-correlation purposes.

\subsection{H0557-383}

This is one of the most luminous galaxies in the sample and it shows clear
evidence for a delay of about 305d between variations at $J$ and $L$ (See
also Fig 1). 

\subsection{F265}

$U$ points from Winkler et al (1992) and Winkler (1997) are shown.

\subsection{IRAS09149-6206}

This is the most luminous galaxy in the sample but unfortunately there are
insufficient data to determine a lag.

\subsection{NGC\,2992}

This galaxy showed some form of outburst, much more luminous than any normal
supernova, at around JD2447200. The cross-correlation programme described
here does not yield a significant delay between $J$ and $K$. However,
Oknyanskij \& Horne (2001) find 18 $\pm$ 10d between $J$ and $K$ and 84
$\pm$ 30d between $J$ and $L$. The programme used here gets a delay of about
35d between $J$ and $L$.

The $K$ vs $L$ diagram shows some curvature instead of a linear relation.
Glass (1997a) mentioned that the time constant for the decay of the outburst
at $L$ is considerably longer than that for $K$.

\subsection{MCG-5-23-16}

This galaxy shows little variation at $J$ and no significant evidence for
delays.

\subsection{NGC\,3783}

The data now presented stretch about 3000 days beyond those of Glass (1992)
which suggested the existence of a delay of about 80 days between
$U$ and $L$. The data now available now suggest a delay of about 190d. It
should be noted, however, that the sampling is rather inadequately spaced.

\subsection{NGC\,4593}

Oknyanskij \& Horne (2001) conclude from the data of Santos-Ll\'{e}o et al
(1994) that there is a delay of 36 $\pm$ 15d between the $J$ and $L$
variations. The present analysis yields 55d.

\subsection{ESO323-G77}

The nucleus of this bright Seyfert 1 galaxy is partially obscured and
shows visible-light polarization explicable by scattering off dust close to
a torus (Schmid, Appenzeller \& Burch, 2003).   
In the infrared it shows a slow decline over $\sim$5000d, by about 0.2 mag at
$L$ and about 0.1 mag at $J$.

\subsection{MCG-6-30-15}

This well-known x-ray galaxy has shown only modest variations in the IR and
its flux variation gradients cannot be determined very accurately. A delay
of about 34d between $J$ and $L$ is indicated.

\subsection{IC4329A}

This has a partially obscured nucleus, very faint at visible wavelengths
(Winkler et al, 1992). There is a suggestion of a delay of between 33 and
85d between $J$ and $L$.

\subsection{NGC\,5506}

This is a heavily obscured nucleus, with colours of its variable part much
redder than average in $J-H$ and $H-K$. A delay of about 25d between $J$ and
$L$ may be present.

\subsection{ESO 103-G35}

Shows no significant variability.

\subsection{F51}

A high-polarization Seyfert 1 galaxy, similar to ESO\,323-G77. The infrared
colours are not noticeably redder than average. There is a delay of about
15d, probably not significant, between $J$ and $L$. The cross-correlation of $U$ with $L$ does not
yield a sensible result, possibly because of the poorer sampling in $U$. 

\subsection{ESO141-G55}

Cross-correlation between $U$ and $L$ gives 230d, while between $J$ and $L$
a delay of $\sim$160d is found.

\subsection{NGC\,6814}

This is a Seyfert with nearly `ordinary' colours much of the time.
Oknyanskij \& Horne (2001) report an upper limit for the delay between $U$
and $K$ of 15d, based on work by B.O. Nelson. The present data give
$\sim$35d.

\subsection{Mkn 509}

For the interval JD\,2447415--9323, Carone et al (1996) have published
visible-region spectrophotometric data which includes $F_{\lambda}$
(5110\AA) data. It is found that these data, converted to a magnitude scale
and assigned a suitable zero point, fit well with the $U$-band data
obtained at SAAO by Winkler et al (1992), Winkler (1997) and as part of the
service programme. Further data by Doroshenko (1996) is also in good
agreement. A hybrid `$U$' light curve has been constructed from these data
and is shown in Fig 1. 

The $L$ band output appears to be heavily smoothed compared to the $U$,
$J$, $H$ and $K$. The delay between $U$ and $L$ is poorly defined but seems
to be about 60d. However, the $K$ flux lags the $U$ by about 100 days. It is
possible that the geometry of the dust shell smears out the $L$ response. It
is also the case that there was no infrared monitoring just after the UV
outbursts at JD2448100 and 2448800, when the long wavelength emission might
have been expected to peak.

The $K$ vs $L$ diagram of Mkn509 shows an atypical slope for the regression
line which makes the $K-L$ colour of the variable component much bluer (0.49
$\pm$ 0.23) than average. If the two brightest $K$ points are omitted, the
slope is changed significantly and $K-L$ has a more typical value, viz 1.13
$\pm$ 0.19. However, no obvious mistake can be found in the photometric
record.

Carone et al (1996) note that the response of H$_{\beta}$ and He{\small II} to
the continuum variations show lags of $\sim$80 and $\sim$60d respectively;
these delays are unusually long.

\subsection{NGC\,7469}

Oknyanskij \& Horne (2001) have taken the data of Glass (1998) and
cross-correlated it with published and unpublished $U$ photometry (e.g.
Doroshenko, Lyutyi \& Rakhimov, 1989; Chuvaev, Lyutyi \& Doroshenko, 1990),
finding that there are delays of 52$\pm$15d at $K$ and 60$\pm$10d at $L$.
The present data, using only the published $U$ photometry, yield, with lower
significance, a delay between $U$ and $L$ of $\sim$21d and between $J$ and
$L$ of $\sim$40d. The Oknyanskij \& Horne (2001) value should probably be
preferred.

\subsection{MCG-2-58-22}

Four RoSat X-ray observations plotted by Kim \& Boller (2002) covering
JD\,2448000 to 2449500 are consistent with the peak we see in the $U$ band
around JD\,2448500. 

This galaxy declined by over a magnitude in $L$ between JD\,2446800 and
JD\,2450000, with minor recoveries. The $L$ vs $K$ diagram, although
affected by observational scatter, seems to show a flattening out, or change
of slope, as the galaxy becomes faint, in the sense that $K$ is tending
towards a constant value as $L$ declines further. There is some indication
of a similar, but less certain, trend in the $K$ vs $H$ diagram. The
explanation may be that the hottest dust, closest to the nucleus in distance
and propagation time, has already cooled in response to the shutting off of
the ultraviolet component, and a cooler dust zone that contributes to the
$L$ radiation is still cooling because it is further away and takes longer
to respond.

The present data suggest a delay of $\sim$160d between $J$ and $L$.

\subsection{NGC\,7603}

NGC\,7603 has been monitored spectrophotometrically in the visible for 20
years by Kollatschny, Bischoff \& Dietrich (2000). Their continuum flux at
5080\AA ~ is plotted in Fig.\ 1. The general features of the contimuum
variations are also seen in the infrared. Unfortunately, there was no IR
coverage around the continuum peak at $\sim$JD\,2449270. The sharp spike in
continuum emission at JD2448862 does not seem to have affected the IR
observation on JD\,2448874. The $JHK$ fluxes and colours at the time of the
first IR observation (JD\,2443719) are consistent with an ordinary underlying
galaxy. From the Kollatschny et al (2000) data it is probable that the
nuclear activity was then close to its lowest point. 

The cross-correlation programme does not yield a useful output.  

\section{Conclusions}

Thirty-nine of the forty-one Seyfert galaxies in this sample have shown
variability during the observing programme; some of them by over
one mag. These variations reflect changes in the ultraviolet output of the
central engine in the same way as the visible-light continuum and spectral
line observations do; i.e., after a certain time-lag.

Lags are common and most conspicuous at $L$, but difficult to determine
accurately. Several of the lower-luminosity galaxies in the present sample
have been shown to vary and to exhibit probable lags on time scales of tens
of days. These will repay study with more frequent sampling. There is some
suggestion that a decline in the $L$ flux following a drop in the UV from
the central engine may show a longer time constant than its counterpart at
$K$, perhaps because of increased radial extent of the dust zone which
contributes most at this wavelength.

For the most part, there are strong linear correlations between the $J,H,K$
and $L$ fluxes. The colours of the variable components are therefore easy to
determine, provided that the variations have sufficient amplitude and the
photometry is sufficiently accurate. There is a high degree of uniformity
of colours among the variable components which allows an estimate of the
nuclear reddening to be made in heavily reddened cases. It also clear that
the near-infrared emission mechanism is the same in all Seyfert galaxies.

In many cases, the fluxes seen through the measuring aperture can be
separated into variable and underlying galaxy components without recourse to
surface-brightness modelling. In a number of galaxies, a further
non-variable component, probably arising from cool dust, is clearly present
at $L$ and may be modelled.

\begin{table}
\caption{Sample data table, for III Zw 2. See web version for remainder.}
\begin{tabular}{llllll}
MJD  &  $J$  &  $H$  &  $K$  &  $L$  & $L_{\rm err}$ \\
3717 & 13.18 & 12.25 & 11.11 &  9.82 & .12 \\
6695 & 13.61 & 12.55 & 11.41 &  9.86 & .14 \\
6989 & 13.71 & 12.67 & 11.54 & 10.01 & .09 \\
7436 & 13.60 & 12.70 & 11.63 &  9.95 & .11 \\
7437 & 13.55 & 12.72 & 11.61 & 10.28 & .10 \\
7342 & 13.55 & 12.68 & 11.65 &       &     \\
7842 & 13.54 & 12.74 & 11.67 & 10.32 & .16 \\
8132 & 13.67 & 12.78 & 11.68 & 10.19 & .05 \\
8172 & 13.59 & 12.69 & 11.62 &  9.91 & .06 \\
8227 & 13.56 & 12.61 & 11.57 & 10.00 & .13 \\
8529 & 13.79 & 12.86 & 11.67 & 10.43 & .22 \\
8574 & 13.64 & 12.76 & 11.60 & 10.07 & .17 \\
8576 & 13.75 & 12.75 & 11.67 & 10.10 & .18 \\
8874 & 13.75 & 12.84 & 11.67 & 10.13 & .05 \\
9966 & 13.84 & 12.89 & 11.79 & 10.41 & .12 \\
10655& 13.61 & 12.66 & 11.56 & 10.31 & .17 \\
10742& 13.64 & 12.61 & 11.69 & 10.00 & .16 \\
11117& 13.60 & 12.71 & 11.60 & 10.43 & .16 \\
\end{tabular}                
\end{table}

\section{Acknowledgments}

I would like to thank Dr David Kilkenny and Fran\c{c}ois van Wyk and Fred
Marang, the observers on the SAAO 0.5-m service programme, for unpublished
$UBVRI$ photometry. Dr Hartmut Winkler (VISTA University) is thanked for
reading an early version of the ms. Dr Chris Koen (SAAO) has helped by
commenting on statistical and other questions. The referee, Dr Victor
Oknyanskij, is thanked for useful comments.

\clearpage

\begin{table}
\caption{Tol0109-38}
\begin{tabular}{lllll}
MJD  &  $J$  &  $H$  &  $K$  &  $L$  \\
5302 & 11.91 & 10.65 & 9.43 & 7.60 \\
5885 & 11.91 & 10.66 & 9.52 & 7.70 \\
6287 & 11.84 & 10.64 & 9.48 & 7.65 \\
6664 & 11.80 & 10.52 & 9.36 & 7.63 \\
6692 & 11.83 & 10.56 & 9.39 & 7.66 \\
7143 & 11.82 & 10.57 & 9.41 & 7.55 \\
7436 & 11.84 & 10.57 & 9.44 & 7.67 \\
7509 & 11.78 & 10.50 & 9.37 & 7.61 \\
7749 & 11.83 & 10.53 & 9.37 & 7.63 \\
7841 & 11.86 & 10.60 & 9.43 & 7.63 \\
8132 & 11.78 & 10.52 & 9.36 & 7.67 \\
8173 & 11.82 & 10.44 & 9.30 & 7.55 \\
8573 & 11.78 & 10.49 & 9.33 & 7.61 \\
8630 & 11.78 & 10.49 & 9.35 & 7.55 \\
8875 & 11.77 & 10.51 & 9.38 & 7.61 \\
8902 & 11.81 & 10.54 & 9.38 & 7.63 \\
8906 & 11.77 & 10.52 & 9.40 & 7.63 \\
8908 & 11.83 & 10.54 & 9.37 & 7.61 \\
9258 & 11.83 & 10.55 & 9.39 & 7.62 \\
9614 & 11.79 & 10.47 & 9.32 & 7.60 \\
9615 & 11.78 & 10.48 & 9.30 & 7.59 \\
9970 & 11.98 & 10.66 & 9.54 & 7.68 \\
9971 & 11.93 & 10.69 & 9.57 & 7.75 \\
10058& 11.86 & 10.62 & 9.52 & 7.74 \\
10117& 11.86 & 10.57 & 9.48 & 7.72 \\
10304& 11.78 & 10.47 & 9.32 & 7.65 \\
10410& 11.79 & 10.50 & 9.36 & 7.55 \\
10413& 11.84 & 10.53 & 9.38 & 7.69 \\
10655& 11.89 & 10.67 & 9.51 & 7.70 \\
10738& 11.91 & 10.68 & 9.55 & 7.77 \\
11115& 11.85 & 10.68 & 9.64 & 7.95 \\
\end{tabular}                
\end{table}
                                   
\begin{table}
\caption{Fairall 9}
\begin{tabular}{lllll}
MJD  &  $J$  &  $H$  &  $K$  &  $L$  \\
3061 & 11.81 & 11.05 & 10.20 &  8.58 \\       
3510 & 11.80 & 10.64 & 09.73 &  8.40 \\      
3681 & 11.79 & 10.75 & 09.80 &  8.43 \\
3683 & 11.85 & 10.75 & 09.84 &  8.44 \\
4204 & 11.72 & 10.62 & 09.60 &  8.26 \\
4469 & 11.89 & 10.69 & 09.64 &  8.37 \\
5341 & 12.16 & 10.97 & 09.91 &  8.48 \\
\end{tabular}                
\end{table}

\begin{table}
\caption{NGC 526a}
\begin{tabular}{llllll}
MJD  &  $J$  &  $H$  &  $K$  &  $L$  & $L_{\rm err}$ \\
4466 & 12.32 & 11.18 & 10.22 &       &     \\
4482 &       &       & 10.32 &  8.58 & .10 \\
4919 & 12.49 & 11.59 & 10.92 &       &     \\
5889 & 12.54 & 11.71 & 11.09 &  9.95 & .15 \\
5890 & 12.48 & 11.64 & 11.06 &       &     \\
6287 & 12.50 & 11.66 & 11.16 &  0.11 & .20 \\
6478 & 12.44 & 11.55 & 10.95 &       &     \\
\end{tabular}                
\end{table}

\begin{table}
\caption{NGC985}
\begin{tabular}{llllll}
MJD  &  $J$  &  $H$  &  $K$  &  $L$  & $L_{\rm err}$ \\
7839  & 12.70 & 11.80 & 11.19 & 10.23 & .11 \\
9972  & 12.65 & 11.78 & 11.13 & 10.07 & .08 \\
10057 & 12.60 & 11.72 & 11.07 &  9.87 & .13 \\
10118 & 12.53 & 11.65 & 11.05 &  9.95 & .12 \\
10721 & 12.39 & 11.46 & 10.71 &  9.55 & .09 \\
11117 & 12.60 & 11.63 & 10.94 &  9.81 & .06 \\
11601 & 12.62 & 11.73 & 11.02 &  9.71 & .24 \\
\end{tabular}                     
\end{table}

\begin{table}
\caption{ESO198-G24}
\begin{tabular}{llllll}
MJD  &  $J$  &  $H$  &  $K$  &  $L$  & $L_{\rm err}$ \\
7842  & 12.88 & 11.92 & 11.10 &  9.75 & .11  \\
8575  & 12.91 & 12.00 & 11.29 &  9.91 & .14  \\
8576  & 12.96 & 12.03 & 11.26 & 10.25 & .15  \\
8630  & 12.93 & 12.03 & 11.27 &  9.97 & .12  \\
8875  & 12.94 & 12.06 & 11.27 & 10.10 & .06  \\
8902  & 12.95 & 12.12 & 11.36 & 10.15 & .14  \\
8906  & 13.00 & 12.08 & 11.29 & 10.10 & .08  \\
8908  & 13.09 & 12.17 & 11.33 & 10.26 & .10  \\
9255  & 12.83 & 12.03 & 11.24 & 10.09 & .17  \\
9614  & 13.18 & 12.30 & 11.52 & 10.03 & .11  \\
9970  & 13.22 & 12.22 & 11.51 & 10.09 & .12  \\
10053 & 13.16 & 12.29 & 11.51 & 10.14 & .16  \\
10116 & 13.22 & 12.29 & 11.58 & 10.31 & .19  \\
10304 & 13.12 & 12.21 & 11.52 & 10.24 & .10  \\
10413 & 13.37 & 12.25 & 11.51 & 10.25 & .12  \\
10721 & 13.32 & 12.57 & 12.02 & 10.85 & .21  \\
10738 & 13.45 & 12.58 & 11.98 &       &      \\
11117 & 13.08 & 12.13 & 11.37 & 10.42 & .10  \\
\end{tabular}                                
\end{table}

\begin{table}
\caption{NGC1068}
\begin{tabular}{lllll}
MJD  &  $J$  &  $H$  &  $K$  &  $L$  \\
0900  &  0.0  & 8.05  & 7.40  & 5.42 \\
3024  & 9.13  & 8.18  & 7.47  & 5.51 \\
6047  & 9.06  & 8.08  & 7.09  & 4.97 \\ 
6314  & 9.05  & 8.09  & 7.10  & 4.96 \\
7437  & 9.03  & 8.03  & 7.01  & 4.90 \\ 
7842  & 9.07  & 8.08  & 7.03  & 4.88 \\
7877  & 9.07  & 8.06  & 7.01  & 4.82 \\
8173  & 9.02  & 8.03  & 6.97  & 4.82 \\
8529  & 9.00  & 8.00  & 6.92  & 4.77 \\
8575  & 9.05  & 8.02  & 6.93  & 4.78 \\
8634  & 9.08  & 8.02  & 6.94  & 4.75 \\
8875  & 9.04  & 7.99  & 6.90  & 4.74 \\
8902  & 9.06  & 8.02  & 6.91  & 4.73 \\
9615  & 9.04  & 7.99  & 6.87  & 4.66 \\
9889  & 0.0   & 0.0   & 6.89  & 4.72 \\ 
9895  & 9.12  & 8.02  & 6.91  & 4.70 \\ 
9966  & 9.06  & 7.99  & 6.89  & 4.69 \\ 
9971  & 9.09  & 8.01  & 6.89  & 4.71 \\
10057 & 9.06  & 8.03  & 6.91  & 4.70 \\
10114 & 9.07  & 8.04  & 6.93  & 4.69 \\
10302 & 9.03  & 8.00  & 6.92  & 4.72 \\
10410 & 9.04  & 8.04  & 6.95  & 4.75 \\
10655 & 9.06  & 8.04  & 6.97  & 4.78 \\
10863 & 9.09  & 8.10  & 7.07  & 4.80 \\
10868 & 9.08  & 8.08  & 7.02  & 4.84 \\
11115 & 9.00  & 8.01  & 7.04  & 4.84 \\
11599 & 9.10  & 8.07  & 7.05  & 4.82 \\
11600 & 9.05  & 8.06  & 7.03  & 4.81 \\
\end{tabular}                       
\end{table}                         
                                    
\begin{table}
\caption{NGC1566}
\begin{tabular}{llllll}
MJD  &  $J$  &  $H$  &  $K$  &  $L$  & $L_{\rm err}$ \\
1569 & 10.59 &  9.66 &  9.41 &       &       \\
3024 &       &       &  9.48 &       &       \\
3025 & 10.64 &  9.80 &  9.50 &  8.77 &   .28 \\
3718 & 10.59 &  9.78 &  9.53 &       &       \\
3887 &       &  9.70 &       &       &       \\
4466 & 10.65 &  9.81 &  9.53 &       &       \\
5266 &       &  0.0  &  9.51 &       &       \\
\end{tabular}                        
\end{table}                          
                                     
\begin{table}
\caption{3C120}
\begin{tabular}{llllll}
MJD  &  $J$  &  $H$  &  $K$  &  $L$  & $L_{\rm err}$ \\
6049 & 12.32 & 11.30 & 10.28 &  8.86 & .06 \\
7143 & 12.48 & 11.49 & 10.54 &  9.08 & .07 \\
7437 & 12.42 & 11.36 & 10.43 &  9.05 & .05 \\
7580 & 12.59 & 11.53 & 10.49 &  9.18 & .09 \\
7749 & 12.40 & 11.45 & 10.52 &  8.99 & .04 \\
7839 & 12.37 & 11.36 & 10.45 &  9.06 & .06 \\
7968 & 12.46 & 11.43 & 10.45 &  9.10 & .09 \\
8172 & 12.42 & 11.35 & 10.40 &  8.97 & .03 \\
8529 & 12.66 & 11.61 & 10.59 &  8.95 & .04 \\
8575 & 12.58 & 11.60 & 10.59 &  9.05 & .05 \\
8629 & 12.65 & 11.61 & 10.62 &  9.12 & .06 \\
8633 & 12.62 & 11.59 & 10.65 &  9.12 & .05 \\
8634 & 12.64 & 11.60 & 10.66 &  9.13 & .05 \\
8874 & 12.42 & 11.58 & 10.71 &  9.26 & .04 \\
8902 & 12.47 & 11.51 & 10.68 &  9.12 & .05 \\
8908 & 12.39 & 11.50 & 10.64 &  9.22 & .05 \\
9255 & 12.16 & 11.21 & 10.29 &  8.98 & .05 \\
9615 & 12.46 & 11.38 & 10.34 &  8.83 & .04 \\
9970 & 12.36 & 11.32 & 10.37 &  8.97 & .05 \\
10114& 12.38 & 11.36 & 10.39 &  8.76 & .01 \\
10410& 12.34 & 11.32 & 10.40 &  8.89 & .06 \\
10507& 12.26 & 11.25 & 10.32 &  8.86 & .04 \\
10657& 12.21 & 11.18 & 10.21 &  8.79 & .04 \\
10720& 12.26 & 11.17 & 10.18 &  8.82 & .05 \\
10738& 12.30 & 11.18 & 10.23 &  8.69 & .06 \\
10863& 12.36 & 11.29 & 10.29 &  8.74 & .08 \\
10868& 12.36 & 11.31 & 10.33 &  8.85 & .07 \\
11115& 12.36 & 11.32 & 10.32 &  9.08 & .05 \\
11599& 12.50 & 11.41 & 10.45 &       &     \\
11600& 12.46 & 11.44 & 10.44 &  9.02 & .10 \\
\end{tabular}                                
\end{table}                                  
                                             
\begin{table}
\caption{Akn120}
\begin{tabular}{lllll}
MJD  &  $J$  &  $H$  &  $K$  &  $L$  \\
6780 &  11.65 & 10.67 &  9.81 &  8.57      \\
7143 &  11.76 & 10.71 &  9.83 &  8.49      \\
7244 &  11.67 & 10.70 &  9.80 &  8.55      \\
7437 &  11.80 & 10.76 &  9.90 &  8.49      \\
7513 &  11.70 & 10.73 &  9.85 &  8.60      \\
7580 &  11.68 & 10.72 &  9.85 &  8.56      \\
7839 &  11.57 & 10.57 &  9.70 &  8.43      \\
7875 &  11.56 & 10.54 &  9.67 &  8.43      \\
7876 &  11.64 & 10.55 &  9.70 &  8.47      \\
7967 &  11.59 & 10.53 &  9.67 &  8.34      \\
8172 &  11.57 & 10.53 &  9.66 &  8.29      \\
8529 &  11.92 & 10.86 &  9.94 &  8.59      \\
8575 &  11.92 & 10.92 & 10.02 &  8.67      \\
8576 &  11.94 & 10.94 & 10.03 &  8.61      \\
8629 &  11.93 & 10.92 & 10.06 &  8.63      \\
8633 &  11.89 & 10.94 & 10.07 &  8.62      \\
8874 &  11.75 & 10.77 &  9.96 &  8.70      \\
8902 &  11.77 & 10.80 &  9.97 &  8.69      \\
8995 &  11.83 & 10.82 &  9.96 &  8.72      \\
9255 &  11.81 & 10.87 & 10.03 &  8.74      \\
9615 &  11.96 & 10.97 & 10.14 &  8.83      \\
9966 &  11.88 & 10.89 & 10.08 &  8.84      \\
10113&  11.96 & 10.93 & 10.11 &  8.88      \\
10410&  11.85 & 10.86 & 10.09 &  8.90      \\
10507&  11.80 & 10.83 & 10.02 &  8.72      \\
10657&  11.82 & 10.83 & 10.03 &  8.78      \\
10720&  11.80 & 10.85 & 10.02 &  8.71      \\
10738&  11.80 & 10.81 & 10.00 &  8.65      \\
10863&  11.76 & 10.81 & 10.02 &  8.66      \\
10868&  11.75 & 10.79 &  9.97 &  8.76      \\
11115&  11.82 & 10.84 & 10.05 &  8.84      \\
11599&  11.84 & 10.96 & 10.18 &  8.93      \\
\end{tabular}                 
\end{table}

\begin{table}
\caption{MCG-5-13-17}
\begin{tabular}{llllll}
MJD  &  $J$  &  $H$  &  $K$  &  $L$  & $L_{\rm err}$ \\
6923 & 11.67 & 10.96 & 10.61 &  9.89 & .05  \\
7144 & 11.74 & 10.96 & 10.51 &  9.78 & .17  \\
7246 & 11.69 & 10.97 & 10.54 &  9.87 & .10  \\
7437 & 11.70 & 10.92 & 10.57 &  9.89 & .08  \\
7580 & 11.81 & 10.98 & 10.59 &  9.63 & .05  \\
7581 & 11.79 & 10.95 & 10.57 &  9.83 & .08  \\
7634 & 11.76 & 11.00 & 10.60 & 10.11 & .08  \\
7842 & 11.77 & 10.99 & 10.63 & 10.02 & .10  \\
7876 & 11.77 & 10.97 & 10.61 & 10.00 & .10  \\
7964 & 11.80 & 11.04 & 10.70 &  9.95 & .16  \\
8173 & 11.80 & 11.02 & 10.73 & 10.03 & .11  \\
8576 & 11.76 & 10.96 & 10.61 & 10.00 & .10  \\
8629 & 11.78 & 11.02 & 10.70 & 10.09 & .14  \\
8877 & 11.78 & 11.02 & 10.65 & 10.05 & .12  \\
8902 & 11.79 & 11.04 & 10.72 & 10.25 & .10  \\
8908 & 11.79 & 11.03 & 10.72 & 10.07 & .12  \\
9615 & 11.71 & 10.95 & 10.56 &  9.90 & .11  \\
10053& 11.70 & 10.92 & 10.50 &  9.65 & .08  \\
10113& 11.75 & 10.97 & 10.60 &  9.84 & .05  \\
10115& 11.73 & 10.96 & 10.62 &  9.84 & .15  \\
10411& 11.70 & 10.93 & 10.53 &  9.77 & .11  \\
10509& 11.82 & 11.00 & 10.69 & 10.04 & .2   \\
10863& 11.80 & 10.99 & 10.63 &  9.84 & .17  \\
11115& 11.72 & 10.93 & 10.56 &  9.78 & .09  \\
11597& 11.74 & 10.92 & 10.55 &  9.66 & .15  \\
\end{tabular}                                
\end{table}                                   
                                           
\begin{table}                              
\caption{Pictor A}                         
\begin{tabular}{lllll}                    
MJD  &  $J$  &  $H$  &  $K$  &  $L$ \\
3887  & 14.04 & 12.98 & 11.92 &  \\
5743  & 14.07 & 12.94 & 11.89 &  \\
6050  & 13.95 & 12.99 & 12.05 &  \\
6776  &       & 12.93 & 11.97 &  \\
7144  & 14.04 & 12.95 & 11.81 &  \\
7247  & 13.92 & 12.88 & 11.88 &  \\
7584  & 14.12 & 13.17 & 12.13 &  \\
7842  & 13.94 & 12.97 & 12.05 &  \\
7877  & 13.96 & 12.98 & 12.06 &  \\
8579  & 14.21 & 13.05 & 12.14 &  \\
8630  & 13.88 & 12.99 & 12.05 &  \\
9971  &       & 12.86 & 11.89 &  \\
10115 & 14.01 & 12.88 & 12.07 &  \\
10509 & 13.85 & 12.96 & 12.01 &  \\
10600 & 14.18 & 13.06 & 12.09 &  \\
10721 & 13.79 & 12.93 & 11.96 &  \\
10866 & 13.87 & 12.81 & 11.90 & 10.34 \\
\end{tabular}                                
\end{table}                                   
      
\begin{table}
\caption{NGC2110}
\begin{tabular}{lllll}
MJD  &  $J$  &  $H$  &  $K$  &  $L$   \\
4320  & 10.83 &  9.86 &  9.44 &       \\
5266  & 10.79 &  9.88 &  9.50 &       \\
5343  & 10.84 &  9.90 &  9.48 &  8.68 \\
7347  & 10.83 &  9.81 &  9.44 &  8.75 \\
7580  & 10.82 &  9.86 &  9.44 &  8.76 \\
7839  & 10.84 &  9.86 &  9.47 &  8.78 \\
7876  & 10.84 &  9.87 &  9.45 &  8.74 \\
7967  & 10.84 &  9.90 &  9.47 &  8.74 \\
8172  & 10.82 &  9.86 &  9.45 &  8.80 \\
8874  & 10.83 &  9.85 &  9.48 &  8.80 \\
9972  & 10.82 &  9.85 &  9.43 &  8.64 \\
10113 & 10.81 &  9.81 &  9.40 &  8.62 \\
10507 & 10.83 &  9.86 &  9.44 &  8.74 \\
11600 & 10.88 &  9.88 &  9.47 &  8.61 \\
11602 & 10.79 &  9.83 &  9.44 &  8.52 \\
\end{tabular}                                  
\end{table}                                   

\begin{table}
\caption{H0557-383}
\begin{tabular}{lllll}
MJD  &  $J$  &  $H$  &  $K$  &  $L$   \\
4974 &  12.23 & 11.22 & 10.05 &  8.28 \\
5042 &  12.28 & 11.16 & 10.05 &  8.13 \\
6050 &  12.47 & 11.33 & 10.18 &  8.32 \\
6118 &  12.43 & 11.33 & 10.18 &  8.40 \\
6780 &  12.27 & 11.16 &  9.99 &  8.27 \\
6925 &  12.25 & 11.12 &  9.99 &  8.26 \\
7243 &  12.37 & 11.24 & 10.09 &  8.24 \\
7347 &  12.34 & 11.15 & 10.05 &  8.27 \\
7579 &  12.41 & 11.21 & 10.08 &  8.31 \\
7580 &  12.41 & 11.21 & 10.09 &  8.22 \\
7634 &  12.37 & 11.15 & 10.06 &  8.24 \\
7839 &  12.39 & 11.21 & 10.06 &  8.18 \\
7876 &  12.43 & 11.26 & 10.10 &  8.30 \\
7967 &  12.49 & 11.30 & 10.18 &  8.24 \\
8173 &  12.53 & 11.35 & 10.23 &  8.35 \\
8576 &  12.39 & 11.23 & 10.13 &  8.38 \\
8629 &  12.35 & 11.17 & 10.11 &  8.31 \\
8633 &  12.29 & 11.17 & 10.11 &  8.37 \\
8634 &  12.30 & 11.18 & 10.09 &  8.28 \\
8734 &  12.29 & 11.10 & 10.04 &  8.27 \\
8875 &  12.32 & 11.12 & 10.00 &  8.25 \\
8902 &  12.27 & 11.13 &  9.98 &  8.26 \\
9255 &  12.41 & 11.28 & 10.11 &  8.30 \\
9615 &  12.48 & 11.36 & 10.23 &  8.37 \\
9971 &  12.33 & 11.16 & 10.05 &  8.44 \\
10053&  12.21 & 11.11 & 10.01 &  8.25 \\
10113&  12.25 & 11.05 &  9.95 &  8.30 \\
10116&  12.21 & 11.05 &  9.96 &  8.30 \\
10411&  12.20 & 11.06 &  9.93 &  8.16 \\
10509&  12.18 & 11.05 &  9.93 &  8.10 \\
10742&  12.27 & 11.11 & 10.01 &  8.27 \\
11115&  12.24 & 11.03 &  9.86 &  8.14 \\
11597&  12.41 & 11.22 & 10.09 &  8.25 \\
\end{tabular}                                
\end{table}                                   

\begin{table}
\caption{Fairall 265}
\begin{tabular}{llllll}
MJD  &  $J$  &  $H$  &  $K$  &  $L$  & $L_{\rm err}$ \\
5341  & 12.71 & 11.84 & 11.14 &       &     \\
6923  & 12.71 & 11.81 & 11.15 &  9.89 &  .17\\
7243  & 12.70 & 11.83 & 11.12 & 10.00 &  .10\\
7581  & 12.71 & 11.83 & 11.16 & 10.03 &  .07\\
7634  & 12.79 & 11.88 & 11.18 &       &     \\
7967  & 12.73 & 11.88 & 11.17 & 10.34 &  .16\\
8629  & 12.76 & 11.84 & 11.21 & 10.21 &  .14\\
8734  & 12.68 & 11.78 & 11.12 & 10.22 &  .12\\
10114 & 12.79 & 11.84 & 11.17 &  9.97 &  .16\\
10116 & 12.74 & 11.84 & 11.17 & 10.16 &  .18\\
10511 & 12.86 & 11.94 & 11.34 & 10.15 &  .20\\
10868 & 12.75 & 11.86 & 11.23 & 10.15 &  .13\\
\end{tabular}                               
\end{table}                                  

\begin{table}
\caption{IRAS09149-6206}
\begin{tabular}{lllll}
MJD  &  $J$  &  $H$  &  $K$  &  $L$   \\
10114 & 11.71 & 10.61 &  9.46 &  7.90  \\
10116 & 11.70 & 10.64 &  9.51 &  7.90 \\
10213 & 11.68 & 10.60 &  9.50 &  7.93 \\
10510 & 11.58 & 10.52 &  9.42 &  7.88 \\
10867 & 11.75 & 10.37 &  9.25 &  7.72 \\
10950 & 11.51 & 10.42 &  9.34 &  7.83 \\
11597 & 11.74 & 10.57 &  9.39 &  7.90 \\
\end{tabular}                                
\end{table}                                   

\begin{table}
\caption{NGC2992}
\begin{tabular}{llllll}
MJD  &  $J$  &  $H$  &  $K$  &  $L$  & $L_{\rm err}$ \\
3596 & 11.02 & 10.06 &  9.40 &  8.30 & .15 \\          
3681 & 11.15 & 10.17 &  9.50 &  8.38 & .05 \\
4318 & 11.16 & 10.16 &  9.63 &       &     \\
4920 & 11.20 & 10.29 &  9.80 &       &     \\      
5082 & 11.07 & 10.15 &  9.60 &  8.42 & .28 \\
5340 & 11.20 & 10.23 &  9.58 &  8.69 & .05 \\
5740 & 11.18 & 10.16 &  9.63 &  8.60 & .07 \\
\end{tabular}                              
\end{table}                           
                                      
\clearpage

\begin{table}                         
\caption{MCG-5-23-16}                 
\begin{tabular}{llllll}               
MJD  &  $J$  &  $H$  &  $K$  &  $L$  & $L_{\rm err}$ \\
3887  & 11.48 & 10.52 &  9.86 &  8.40 & 0.12   \\
4001  & 11.40 & 10.48 &  9.85 &  8.44 & 0.12   \\
5046  & 11.29 & 10.45 &  9.88 &       &        \\
6612  & 11.33 & 10.43 &  9.84 &  8.63 &  0.05  \\
6886  & 11.39 & 10.53 & 10.03 &       &        \\
6923  & 11.34 & 10.50 &  9.95 &  8.62 &  0.04  \\
6985  & 11.45 & 10.56 & 10.05 &  8.89 &  0.05  \\
7246  & 11.32 & 10.41 &  9.87 &  8.65 &  0.05  \\
7343  & 11.45 & 10.49 &  9.93 &  8.88 &  0.07  \\
7581  & 11.42 & 10.58 & 10.13 &  9.03 &  0.03  \\
7634  & 11.44 & 10.55 & 10.11 &  9.06 &  0.04  \\
7876  & 11.40 & 10.52 & 10.05 &  9.03 &  0.08  \\
7968  & 11.41 & 10.55 & 10.10 &  9.09 &  0.06  \\
8629  & 11.39 & 10.51 & 10.01 &  8.84 &  0.06  \\
8634  & 11.40 & 10.51 & 10.02 &  8.88 &  0.04  \\
8671  & 11.39 & 10.52 & 10.01 &  9.04 &  0.07  \\
8731  & 11.38 & 10.54 & 10.09 &       &        \\
8733  & 11.39 & 10.51 & 10.08 &  9.11 &  0.05  \\
8734  & 11.39 & 10.50 & 10.08 &  8.95 &  0.06  \\
10113 & 11.33 & 10.37 &  9.76 &  8.60 &  0.04  \\
10213 & 11.32 & 10.40 &  9.82 &  8.65 &  0.03  \\
10509 & 11.35 & 10.44 &  9.88 &  8.65 &  0.06  \\
10866 & 11.35 & 10.40 &  9.81 &  8.45 &  0.04  \\
10949 & 11.36 & 10.42 &  9.82 &  8.56 &  0.03  \\
11597 & 11.35 & 10.46 &  9.90 &  8.84 &  0.08  \\
\end{tabular}                                  
\end{table}                                   
      
\begin{table}
\caption{NGC3783}
\begin{tabular}{lllll}
MJD  &  $J$  &  $H$  &  $K$  &  $L$   \\
2225 &        &       &  9.55 &       \\
2822 &        &       &  9.71 &  8.69 \\
3322 &  11.41 & 10.54 &  9.60 &       \\
3511 &        &       &  9.82 &       \\
3602 &  11.54 & 10.67 & 10.09 &       \\
3680 &        &       & 10.26 &       \\
3888 &        &       & 10.00 &       \\
\end{tabular}                                
\end{table}                                   

\begin{table}
\caption{NGC3783 (contd.)}
\begin{tabular}{lllll}
MJD  &  $J$  &  $H$  &  $K$  &  $L$   \\
8634 &  11.34 & 10.45 &  9.71 &  8.37 \\
8671 &  11.35 & 10.46 &  9.69 &  8.34 \\
8674 &  11.32 & 10.42 &  9.67 &  8.29 \\
8675 &  11.32 & 10.41 &  9.65 &  8.23 \\
8700 &  11.40 & 10.45 &  9.70 &  8.39 \\
8727 &  11.33 & 10.43 &  9.69 &  8.34 \\
8731 &  11.34 & 10.45 &  9.72 &  8.31 \\
8733 &  11.35 & 10.43 &  9.69 &  8.38 \\
8734 &  11.33 & 10.45 &  9.70 &  8.39 \\
8765 &  11.24 & 10.44 &  9.72 &  8.30 \\
8768 &  11.28 & 10.47 &  9.72 &  8.36 \\
8783 &  11.17 & 10.39 &  9.65 &  8.31 \\
8786 &  11.09 & 10.27 &  9.62 &  8.32 \\
8789 &  11.16 & 10.34 &  9.60 &  8.32 \\
8790 &  11.07 & 10.27 &  9.58 &  8.24 \\
8792 &  11.16 & 10.35 &  9.63 &  8.27 \\
8804 &  11.13 & 10.29 &  9.59 &  8.22 \\
8834 &  11.20 & 10.35 &  9.60 &  8.26 \\
8993 &  11.22 & 10.21 &  9.36 &  8.00 \\
9890 &  11.58 & 10.68 &  9.98 &  8.57 \\
10113&  11.48 & 10.53 &  9.82 &  8.45 \\
10117&  11.41 & 10.58 &  9.86 &  8.54 \\
10214&  11.46 & 10.59 &  9.93 &  8.58 \\
10304&  11.39 & 10.49 &  9.80 &  8.47 \\
10411&  11.33 & 10.44 &  9.75 &  8.53 \\
10509&  11.32 & 10.34 &  9.58 &  8.25 \\
10615&  11.35 & 10.36 &  9.55 &  8.16 \\
10655&  11.42 & 10.45 &  9.66 &  8.26 \\
10657&  11.42 & 10.45 &  9.64 &  8.19 \\
10863&  11.43 & 10.57 &  9.90 &  8.42 \\
10948&  11.40 & 10.48 &  9.79 &  8.38 \\
11597&  11.38 & 10.45 &  9.71 &  8.51 \\
11600&  11.31 & 10.46 &  9.94 &  8.46 \\
\end{tabular}                                
\end{table}                                   

\begin{table}
\caption{H1143-182}
\begin{tabular}{llllll}
MJD  &  $J$  &  $H$  &  $K$  &  $L$  & $L_{\rm err}$ \\
6611  & 12.84 & 12.00 & 11.33 & 10.33 & .10 \\
6582  & 12.90 & 12.00 & 11.36 &       &     \\
6847  &       &       & 11.35 & 10.43 & .19 \\
7246  & 12.96 & 12.06 & 11.36 &       &     \\
7581  & 12.91 & 12.10 & 11.41 & 10.36 & .07 \\
7634  & 12.94 & 12.04 & 11.35 & 10.63 & .00 \\
7967  & 13.00 & 12.15 & 11.42 & 10.14 & .11 \\
7968  & 12.99 & 12.15 & 11.40 & 10.23 & .12 \\
8042  & 12.87 & 12.12 & 11.38 & 10.27 & .09 \\
8633  & 12.73 & 11.90 & 11.13 & 10.01 & .11 \\
10114 & 13.43 & 12.47 & 11.77 & 10.44 & .23 \\
10214 & 13.40 & 12.55 & 11.79 &       &     \\
10510 & 13.13 & 12.34 & 11.64 & 10.72 & .17 \\
10615 & 13.32 & 12.39 & 11.63 &       &     \\
11599 & 13.10 & 12.19 & 11.54 & 10.00 & .16 \\
\end{tabular}                                
\end{table}                                   
                                   
\begin{table}                      
\caption{I Zw 96}                  
\begin{tabular}{llll}            
MJD  &  $J$  &  $H$  &  $K$   \\
5083  & 13.18 & 12.39 & 11.91 \\           
5741  & 13.16 & 12.48 & 11.92 \\           
5846  & 13.14 & 12.34 & 11.88 \\           
5887  & 13.14 & 12.36 & 11.96 \\
6120  & 13.16 & 12.42 & 11.95 \\
6477  & 13.06 & 12.26 & 11.80 \\
6578  & 13.00 & 12.25 & 11.81 \\
6610  & 13.06 & 12.23 & 11.81 \\
6923  & 13.04 & 12.22 & 11.77 \\
7246  & 13.00 & 12.25 & 11.78 \\
7634  & 13.09 & 12.20 & 11.81 \\
10116 & 13.07 & 12.24 & 11.81 \\
10510 & 13.12 & 12.25 & 11.85 \\
\end{tabular}                                
\end{table}                                    

\begin{table}
\caption{NGC4593}
\begin{tabular}{llllll}
MJD  &  $J$  &  $H$  &  $K$  &  $L$   \\
5081 &  11.08 & 10.28 & 09.80 &       \\
5419 &  11.20 & 10.36 & 09.87 & 08.90 \\
6116 &  11.14 & 10.24 & 09.69 & 08.69 \\
6117 &  11.15 & 10.27 & 09.71 & 08.88 \\
6120 &  11.12 & 10.26 & 09.75 & 08.75 \\
6122 &  11.15 & 10.28 & 09.76 & 08.81 \\
6111 &  11.12 & 10.22 & 09.63 & 08.71 \\
6112 &  11.11 & 10.23 & 09.67 & 08.61 \\
6114 &  11.14 & 10.26 & 09.70 & 08.78 \\
6194 &  11.22 & 10.43 & 09.92 & 08.99 \\
6195 &  11.19 & 10.33 & 09.78 & 08.90 \\
6199 &  11.22 & 10.37 & 09.89 & 09.06 \\
6477 &  11.11 & 10.26 & 09.73 & 08.88 \\
6478 &  11.10 & 10.23 & 09.72 & 08.84 \\
6479 &  11.12 & 10.25 & 09.74 & 08.88 \\
6582 &  11.05 & 10.12 & 09.54 & 08.54 \\
6609 &  11.06 & 10.16 & 09.56 & 08.73 \\
6845 &  11.08 & 10.18 & 09.68 & 08.51 \\
6923 &  11.11 & 10.24 & 09.68 & 08.61 \\
6985 &  11.07 & 10.20 & 09.67 & 08.60 \\
6988 &  11.08 & 10.24 & 09.69 & 08.71 \\
7246 &  11.05 & 10.16 & 09.59 & 08.52 \\
7584 &  11.21 & 10.33 & 09.87 & 08.95 \\
7634 &  11.20 & 10.32 & 09.89 & 08.99 \\
7967 &  11.21 & 10.35 & 09.93 & 09.03 \\
7968 &  11.20 & 10.35 & 09.92 & 09.25 \\
8042 &  11.22 & 10.33 & 09.91 & 09.14 \\
8630 &  11.09 & 10.20 & 09.65 & 08.69 \\
8634 &  11.10 & 10.18 & 09.64 & 08.78 \\
8671 &  11.14 & 10.23 & 09.70 & 08.82 \\
8674 &  11.18 & 10.22 & 09.70 & 08.79 \\
8727 &  11.13 & 10.23 &  9.77 &  8.82 \\
8731 &  11.10 & 10.22 &  9.72 &  8.83 \\
8733 &  11.16 & 10.23 &  9.73 &  8.78 \\
8734 &  11.10 & 10.22 &  9.73 &  8.86 \\
9890 &  11.12 & 10.20 &  9.68 &  8.77 \\
10114&  11.11 & 10.17 &  9.62 &  8.57 \\
10509&  11.10 & 10.13 &  9.53 &  8.38 \\
10615&  11.05 & 10.15 &  9.65 &  8.58 \\
10657&  11.09 & 10.18 &  9.64 &  8.53 \\
10948&  11.10 & 10.19 & 9.63  & 8.59 \\
\end{tabular}                                
\end{table}                                   

\begin{table}
\caption{NGC4748}
\begin{tabular}{llllll}
MJD  &  $J$  &  $H$  &  $K$  &  $L$  & $L_{\rm err}$ \\
6194  & 12.05 & 11.25 & 10.76 & 10.15 & 0.21  \\
6197  & 12.05 & 11.26 & 10.76 & 10.05 & 0.09  \\
6923  & 12.16 & 11.36 & 10.98 & 10.37 & 0.12  \\
6984  & 12.16 & 11.38 & 10.98 & 10.46 & 0.21  \\
7246  & 12.12 & 11.35 & 10.91 &       &       \\
7579  & 12.17 & 11.36 & 10.98 &       &       \\
7580  & 12.17 & 11.39 & 11.02 & 10.54 & 0.14  \\
7634  & 12.15 & 11.31 & 10.94 & 10.51 & 0.11  \\
7967  & 12.20 & 11.35 & 10.99 & 11.44 & 0.11  \\
8042  & 12.20 & 11.36 & 11.02 & 10.43 & 0.10  \\
8674  & 12.21 & 11.39 & 11.03 &       &       \\
8734  & 12.19 & 11.41 & 11.04 & 10.79 & 0.30  \\
10115 & 12.21 & 11.37 & 11.02 & 10.60 & 0.26  \\
10510 & 12.17 & 11.35 & 10.97 & 10.58 & 0.15  \\
\end{tabular}                                
\end{table}                                   
                                    
\begin{table}
\caption{ESO323-G77}
\begin{tabular}{llllll}
MJD  &  $J$  &  $H$  &  $K$  &  $L$  & $L_{\rm err}$ \\
6886  & 10.80 &  9.64 &  8.75 &  7.50 \\
6923  & 10.79 &  9.63 &  8.75 &  7.49 \\
6984  & 10.80 &  9.61 &  8.70 &  7.45 \\
6989  & 10.88 &  9.67 &  8.79 &  7.57 \\
7245  & 10.80 &  9.67 &  8.76 &  7.57 \\
7246  & 10.83 &  9.70 &  8.80 &  7.56 \\
7342  & 10.88 &  9.76 &  8.85 &  7.62 \\
7580  & 10.88 &  9.74 &  8.86 &  7.56 \\
7634  & 10.89 &  9.72 &  8.84 &  7.56 \\
7967  & 10.94 &  9.83 &  8.96 &  7.75 \\
8042  & 10.96 &  9.84 &  8.96 &  7.75 \\
8629  & 11.00 &  9.88 &  9.03 &  7.75 \\
8633  & 11.00 &  9.89 &  9.04 &  7.73 \\
8674  & 10.99 &  9.89 &  9.04 &  7.72 \\
8675  & 10.99 &  9.89 &  9.03 &  7.77 \\
8731  & 10.98 &  9.86 &  9.01 &  7.78 \\
8733  & 10.93 &  9.81 &  8.94 &  7.71 \\
8734  & 10.99 &  9.85 &  8.98 &  7.78 \\
9889  & 11.05 &  9.96 &  9.14 &  7.94 \\
10114 & 11.02 &  9.92 &  9.10 &  7.87 \\
10304 & 11.07 &  9.98 &  9.16 &  8.00 \\
10509 & 11.04 &  9.94 &  9.13 &  7.94 \\
10614 & 11.08 &  9.97 &  9.14 &  7.96 \\
10656 & 11.07 &  9.98 &  9.17 &  7.95 \\
10863 & 11.09 &  10.02&  9.19 &  7.83 \\
10948 & 11.10 &  10.03&  9.18 &  8.03 \\
11599 & 11.00 &  9.96 &  9.18 &  7.93 \\
\end{tabular}                                
\end{table}                                   

\begin{table}
\caption{MCG-6-30-15}
\begin{tabular}{llllll}
MJD  &  $J$  &  $H$  &  $K$  &  $L$  & $L_{\rm err}$ \\
3887  & 11.60 & 10.65 &  9.87 &  8.52 & .10 \\ 
4318  & 11.42 & 10.50 &  9.75 &       &     \\
4720  & 11.36 & 10.47 &  9.84 &       &     \\
5849  & 11.50 & 10.62 &  9.97 &  8.64 & .05 \\
5887  & 11.51 & 10.66 & 10.00 &  8.76 & .05 \\
5889  & 11.53 & 10.66 &  9.98 &  8.78 & .07 \\
6117  & 11.40 & 10.51 &  9.83 &  8.59 & .05 \\
6194  & 11.42 & 10.55 &  9.87 &  8.71 & .04 \\
6195  & 11.42 & 10.51 &  9.88 &  8.56 & .05 \\
6476  & 11.46 & 10.53 &  9.82 &  8.50 & .06 \\
6478  & 11.50 & 10.54 &  9.85 &  8.59 & .08 \\
6582  & 11.37 & 10.52 &  9.90 &  8.60 & .03 \\
6610  & 11.43 & 10.55 &  9.91 &  8.59 & .06 \\
6845  & 11.51 & 10.61 &  9.95 &  8.72 & .05 \\
6848  & 11.48 & 10.60 &  9.97 &  8.73 & .06 \\
6923  & 11.45 & 10.60 &  9.97 &  8.74 & .05 \\
6984  & 11.40 & 10.53 &  9.92 &  8.67 & .05 \\
7243  & 11.43 & 10.50 &  9.84 &  8.60 & .03 \\
7581  & 11.48 & 10.59 &  9.88 &  8.60 & .04 \\
7967  & 11.44 & 10.56 &  9.87 &  8.60 & .04 \\
8042  & 11.55 & 10.60 &  9.96 &  8.70 & .04 \\
8731  & 11.51 & 10.55 &  9.91 &  8.73 & .04 \\
10117 & 11.52 & 10.53 &  9.84 &  8.70 & .03 \\
10510 & 11.48 & 10.50 &  9.85 &  8.65 & .04 \\
10948 & 11.39 & 10.55 &  9.87 &  8.57 & .02 \\
\end{tabular}                                
\end{table}                                   

\begin{table}
\caption{IC4329a}
\begin{tabular}{llllll}
MJD  &  $J$  &  $H$  &  $K$  &  $L$  & $L_{\rm err}$ \\
3325 &  11.14 &10.22 &  9.38 &  7.93 \\
3602 &  11.12 &10.27 &  9.38 &  7.80 \\
3682 &  11.16 &10.19 &  9.32 &  7.90 \\
4320 &  11.00 &10.02 &  9.12 &       \\
4323 &        &      &  9.29 &  7.57 \\
4720 &  11.02 &10.06 &  9.16 &       \\
5849 &  10.94 & 9.97 &  9.09 &  7.64 \\
5887 &  10.93 & 9.94 &  9.04 &  7.59 \\
5889 &  10.94 & 9.95 &  9.05 &  7.65 \\
6194 &  11.06 &10.00 &  9.10 &       \\
6195 &  10.96 & 9.99 &  9.12 &  7.61 \\
6216 &  10.95 &10.00 &  9.10 &  7.55 \\
6607 &  11.04 &10.03 &  9.13 &  7.56 \\
6845 &  11.02 &10.03 &  9.14 &  7.72 \\
6849 &  11.07 &10.03 &  9.15 &  7.66 \\
6923 &  11.09 &10.11 &  9.23 &  7.77 \\
6925 &  11.07 &10.10 &  9.22 &  7.68 \\
6984 &  11.11 &10.14 &  9.26 &  7.80 \\
7243 &  11.06 &10.13 &  9.29 &  7.80 \\
7342 &  10.99 &10.06 &  9.24 &  7.77 \\
7581 &  11.05 &10.03 &  9.14 &  7.68 \\
7584 &  11.07 &10.04 &  9.15 &  7.67 \\
7634 &  11.05 &10.02 &  9.15 &  7.73 \\
7967 &  10.94 & 9.92 &  9.04 &  7.54 \\
8042 &  10.94 & 9.95 &  9.05 &  7.62 \\
8630 &  10.88 & 9.84 &  8.92 &  7.39 \\
8632 &  10.91 & 9.86 &  8.93 &  7.48 \\
8633 &  10.90 & 9.85 &  8.93 &  7.39 \\
8671 &  10.87 & 9.85 &  8.93 &  7.43 \\
8731 &  10.87 & 9.84 &  8.92 &  7.47 \\
8734 &  10.89 & 9.83 &  8.93 &  7.42 \\
9889 &  10.95 &10.02 &  9.14 &  7.63 \\
10114&  11.01 &10.04 &  9.21 &  7.60 \\
10304&  10.91 & 9.92 &  9.05 &  7.60 \\
10509&  11.07 &10.04 &  9.16 &  7.59 \\
10614&  11.03 &10.04 &  9.12 &  7.61 \\
10656&  10.97 &10.00 &  9.10 &  7.60 \\
10948&  10.97 & 9.94 &  9.04 &  7.54 \\
11600&  10.93 & 9.94 &  9.05 &  7.56 \\
\end{tabular}                                
\end{table}                                   
                              
\begin{table}                 
\caption{NGC5506}             
\begin{tabular}{llllll}       
MJD  &  $J$  &  $H$  &  $K$  &  $L$  & $L_{\rm err}$ \\
3523 &        & 10.33 &  9.08 &  7.37 \\
3525 &  11.69 & 10.30 &  9.06 &  7.35 \\
5104 &  11.51 &  9.99 &  8.68 &  7.13 \\
5846 &  11.67 & 10.23 &  8.97 &  7.26 \\
5886 &  11.64 & 10.19 &  8.96 &  7.33 \\
5890 &  11.62 & 10.21 &  8.96 &  7.32 \\
6116 &  11.42 &  9.89 &  8.62 &  7.05 \\
6117 &  11.43 &  9.92 &  8.66 &  7.08 \\
6122 &  11.45 &  9.92 &  8.68 &  7.06 \\
6107 &  11.34 &  9.81 &  8.56 &  7.05 \\
6110 &  11.39 &  9.87 &  8.62 &  7.17 \\
6112 &  11.40 &  9.87 &  8.64 &  7.10 \\
6198 &  11.47 &  9.99 &  8.76 &  7.17 \\
6477 &  11.41 &  9.87 &  8.62 &  7.08 \\
6582 &  11.49 &  9.99 &  8.70 &  7.12 \\
6609 &  11.47 &  9.98 &  8.72 &  7.13 \\
6845 &  11.32 &  9.71 &  8.38 &  6.86 \\
6846 &  11.33 &  9.72 &  8.40 &  6.82 \\
6847 &  11.35 &  9.75 &  8.44 &  6.91 \\
6850 &  11.32 &  9.71 &  8.39 &  6.88 \\
6923 &  11.46 &  9.90 &  8.56 &  6.96 \\
6925 &  11.49 &  9.90 &  8.57 &  7.02 \\
6984 &  11.46 &  9.93 &  8.60 &  7.05 \\
6989 &  11.47 &  9.89 &  8.59 &  6.96 \\
7243 &  11.43 &  9.89 &  8.58 &  6.99 \\
7581 &  11.55 &  9.88 &  8.55 &  7.01 \\
7584 &  11.50 &  9.86 &  8.55 &  6.92 \\
7628 &  11.47 &  9.86 &  8.56 &  6.94 \\
7634 &  11.46 &  9.87 &  8.55 &  6.94 \\
7967 &  11.50 &  9.97 &  8.67 &  7.02 \\
8042 &  11.44 & 10.00 &  8.67 &  7.09 \\
8671 &  11.56 & 10.03 &  8.77 &  7.13 \\
8674 &  11.53 & 10.01 &  8.74 &  7.15 \\
8733 &  11.59 &  9.99 &  8.72 &  7.15 \\
8734 &  11.50 &  9.95 &  8.69 &  7.12 \\
9890 &  11.51 & 10.14 &  8.83 &  7.23 \\
9891 &  11.64 & 10.15 &  8.86 &  7.18 \\
10117&  11.61 & 10.12 & 8.82  & 7.16  \\
10213&  11.58 & 10.10 & 8.80  & 7.13  \\
10509&  11.59 & 10.17 & 8.93  & 7.33  \\
10615&  11.62 & 10.17 & 8.94  & 7.26  \\
10657&  11.63 & 10.14 & 8.89  & 7.29  \\
10949&  11.65 & 10.18 & 8.97  & 7.32  \\
11599&  11.46 &  9.96 & 8.74  & 7.22  \\
\end{tabular}                                
\end{table}                                   

\begin{table}
\caption{NGC5548}
\begin{tabular}{llllll}
MJD  &  $J$  &  $H$  &  $K$  &  $L$  & $L_{\rm err}$ \\
7243 & 11.62 & 10.81 & 10.10 &  8.91  \\
7246 & 11.62 & 10.83 & 10.10 &  8.83  \\
7241 & 11.61 & 10.79 & 10.08 &  8.67  \\
7248 & 11.59 & 10.78 & 10.08 &  8.74  \\
7342 & 11.66 & 10.80 & 10.06 &  8.65  \\
7581 & 11.67 & 10.81 & 10.06 &  8.79  \\
7584 & 11.69 & 10.76 & 10.02 &  8.89  \\
7628 & 11.59 & 10.71 &  9.99 &  8.65  \\
7634 & 11.63 & 10.71 &  9.98 &  8.84  \\
7748 & 11.67 & 10.71 &  9.94 &  8.56  \\
7967 & 11.81 & 10.88 & 10.13 &  8.77  \\
8042 & 11.79 & 10.96 & 10.29 &  9.00  \\
\end{tabular}                                
\end{table}                                   

\begin{table}
\caption{IRAS15091-2107}
\begin{tabular}{llllll}
MJD  &  $J$  &  $H$  &  $K$  &  $L$  & $L_{\rm err}$ \\
6194  & 12.62 & 11.61 & 10.75 &  9.40 & .07 \\
6923  & 12.65 & 11.68 & 10.79 &  9.43 & .06 \\
6984  & 12.75 & 11.81 & 10.86 &  9.42 & .06 \\
7247  & 12.68 & 11.73 & 10.92 &  9.57 & .08 \\
7585  & 12.77 & 11.82 & 11.02 &       &     \\
7634  & 12.66 & 11.74 & 10.91 &  9.45 & .08 \\
8733  & 12.74 & 11.68 & 10.78 &  9.42 & .09 \\
9891  & 12.51 & 11.66 & 10.95 &  9.60 & .07 \\
10509 & 12.58 & 11.57 & 10.75 &       &     \\
10656 & 12.54 & 11.62 & 10.75 &  9.54 & .07  \\
\end{tabular}                                
\end{table}                                  

\begin{table}
\caption{ESO103-G35}
\begin{tabular}{llllll}
MJD  &  $J$  &  $H$  &  $K$  &  $L$  & $L_{\rm err}$ \\
4320  & 12.37 & 11.50 & 11.07 &       &     \\
5844  & 12.22 & 11.43 & 11.03 &  9.73 & .11 \\
6197  & 12.25 & 11.46 & 11.04 &  9.68 & .08 \\
6608  & 12.25 & 11.47 & 11.09 &       &     \\
6691  & 12.33 & 11.43 & 11.05 &  9.55 & .13 \\
6923  & 12.25 & 11.43 & 11.03 &  9.70 & .09 \\
6984  & 12.33 & 11.50 & 11.10 &  9.73 & .07 \\
7243  & 12.26 & 11.40 & 11.05 &  9.75 & .13 \\
7343  & 12.27 & 11.49 & 11.05 &  9.77 & .13 \\
7433  & 12.30 & 11.44 & 11.06 &  9.78 & .05 \\
7749  & 12.24 & 11.44 & 11.05 &  9.72 & .09 \\
8042  & 12.25 & 11.43 & 11.02 &  9.73 & .09 \\
8132  & 12.23 & 11.41 & 11.05 &  9.74 & .07 \\
8167  & 12.24 & 11.43 & 11.06 &  9.66 & .05 \\
8173  &  0.0  &  0.0  & 11.06 &  9.55 & .09 \\
8528  & 12.25 & 11.45 & 11.07 &  9.88 & .13 \\
8731  & 12.28 & 11.45 & 11.06 &  9.63 & .13 \\
8733  & 12.31 & 11.44 & 11.08 &  9.73 & .08 \\
9615  & 12.26 & 11.42 & 11.04 &  9.64 & .10 \\
9889  & 12.21 & 11.42 & 11.02 &  9.75 & .10 \\
9966  & 12.27 & 11.44 & 11.02 &  9.76 & .10 \\
10212 & 12.27 & 11.42 & 11.05 &  9.76 & .11 \\
10304 & 12.26 & 11.40 & 11.02 &  9.80 & .10 \\
10615 & 12.30 & 11.42 & 11.03 &  9.79 & .10 \\
10656 & 12.23 & 11.38 & 11.01 &  9.78 & .09 \\
10948 & 12.27 & 11.43 & 11.05 &  9.63 & .06 \\
\end{tabular}                                
\end{table}                                   
                                  
\begin{table}
\caption{Fairall 51}
\begin{tabular}{llllll}
MJD  &  $J$  &  $H$  &  $K$  &  $L$  & $L_{\rm err}$ \\
3029  & 12.01 & 10.88 &  9.92 &  8.45 & .09 \\
4320  & 12.10 & 11.00 & 10.07 &       &     \\
5595  & 12.02 & 11.00 & 10.20 &  8.70 & .03 \\
5887  & 11.99 & 11.01 & 10.12 &  8.79 & .05 \\
6923  & 12.04 & 11.09 & 10.29 &  8.95 & .06 \\
6984  & 12.07 & 11.13 & 10.29 &  9.07 & .03 \\
7243  & 12.00 & 11.08 & 10.26 &  8.97 & .07 \\
7342  & 11.99 & 11.01 & 10.17 &  8.85 & .06 \\
7343  & 11.98 & 10.99 & 10.22 &  8.81 & .05 \\
7433  & 11.98 & 10.97 & 10.11 &  8.78 & .05 \\
7749  & 11.94 & 10.90 & 10.04 &  8.68 & .03 \\
7842  & 12.09 & 11.10 & 10.25 &  8.83 & .06 \\
7967  & 11.88 & 10.81 &  9.96 &  8.63 & .06 \\
8042  & 11.91 & 10.87 &  9.98 &  8.60 & .03 \\
8132  & 11.95 & 10.88 & 10.01 &  8.65 & .03 \\
8167  & 11.93 & 10.88 &  9.98 &  8.67 & .04 \\
8528  & 11.95 & 10.92 & 10.06 &  8.72 & .05 \\
8731  & 12.02 & 10.95 & 10.04 &  8.75 & .04 \\
8733  & 11.97 & 10.92 & 10.03 &  8.71 & .04 \\
8734  & 11.97 & 10.91 & 10.01 &  8.69 & .06 \\
8875  & 12.00 & 10.99 & 10.14 &  8.84 & .03 \\
9889  & 12.14 & 11.04 & 10.17 &  8.80 & .06 \\
9966  & 12.17 & 11.14 & 10.26 &  8.75 & .05 \\
9972  & 12.15 & 11.12 & 10.28 &  8.89 & .05 \\
10213 & 12.06 & 10.99 & 10.16 &  8.80 & .05 \\
10304 & 11.94 & 10.86 &  9.99 &  8.77 & .04 \\
10614 & 11.94 & 10.90 & 10.06 &  8.74 & .04 \\
10656 & 11.96 & 10.90 & 10.04 &  8.76 & .03 \\
10948 & 12.01 & 10.93 & 10.01 &  8.66 & .03\\
\end{tabular}                                
\end{table}                                   

\begin{table}
\caption{ESO141-G55}
\begin{tabular}{llllll}
MJD  &  $J$  &  $H$  &  $K$  &  $L$  & $L_{\rm err}$ \\
4320  & 12.00 & 10.98 & 10.05 &       &     \\
5238  & 12.17 & 11.08 & 10.08 &       &     \\
5596  & 12.42 & 11.42 & 10.42 &  8.94 & .04 \\
5889  & 12.26 & 11.29 & 10.39 &  8.96 & .05 \\
6197  & 12.18 & 11.21 & 10.27 &  8.92 & .02 \\
6664  & 12.41 & 11.38 & 10.41 &  9.00 & .06 \\
6694  & 12.38 & 11.41 & 10.47 &  9.15 & .12 \\
6923  & 12.18 & 11.31 & 10.40 &  9.06 & .07 \\
6984  & 12.27 & 11.26 & 10.36 &  8.97 & .04 \\
7243  & 12.29 & 11.31 & 10.33 &  8.93 & .08 \\
7342  & 12.55 & 11.51 & 10.52 &  9.01 & .07 \\
7343  & 12.53 & 11.47 & 10.57 &  9.10 & .07 \\
7433  & 12.43 & 11.53 & 10.62 &  9.08 & .03 \\
7435  & 12.40 & 11.45 & 10.56 &  9.19 & .06 \\
7748  & 12.21 & 11.37 & 10.50 &  9.16 & .05 \\
7842  & 12.25 & 11.27 & 10.39 &  9.03 & .06 \\
7967  & 12.36 & 11.36 & 10.43 &  9.15 & .07 \\
8042  & 12.30 & 11.38 & 10.48 &  9.09 & .04 \\
8132  & 12.30 & 11.33 & 10.46 &  9.15 & .04 \\
8167  & 12.31 & 11.36 & 10.48 &  9.14 & .05 \\
8528  & 12.35 & 11.36 & 10.44 &  9.04 & .07 \\
8731  & 12.27 & 11.23 & 10.38 &  9.09 & .06 \\
8734  & 12.23 & 11.25 & 10.35 &  9.01 & .07 \\
8874  & 12.22 & 11.23 & 10.33 &  9.02 & .03 \\
9255  & 12.25 & 11.33 & 10.45 &  9.06 & .06 \\
9610  & 12.35 & 11.32 & 10.35 &  9.07 & .04 \\
9889  & 12.59 & 11.56 & 10.64 &  9.15 & .07 \\
9966  & 12.59 & 11.64 & 10.76 &  9.51 & .08 \\
9972  & 12.53 & 11.63 & 10.75 &  9.21 & .05 \\
10212 & 12.40 & 11.53 & 10.68 &  9.42 & .12 \\
10302 & 12.45 & 11.46 & 10.63 &  9.39 & .06 \\
10614 & 12.28 & 11.31 & 10.45 &  9.13 & .03 \\
10655 & 12.19 & 11.27 & 10.44 &  9.17 & .07 \\
10948 & 12.40 & 11.43 & 10.51 &  9.16 & .03 \\
\end{tabular}                                
\end{table}                                   

\begin{table}
\caption{NGC6814}
\begin{tabular}{llllll}
MJD  &  $J$  &  $H$  &  $K$  &  $L$  & $L_{\rm err}$ \\
3024 &  11.53 & 10.47 & 10.14 &       &     \\
3683 &  11.53 & 10.71 & 10.32 &  9.80 & .25 \\
5585 &  11.29 & 10.44 & 10.02 &  9.45 & .00 \\
5888 &  11.37 & 10.47 & 10.06 &       &     \\
6191 &  11.40 & 10.53 & 10.20 &  9.76 & .08 \\
6287 &  11.44 & 10.58 & 10.25 &  9.73 & .13 \\
6329 &  11.41 & 10.58 & 10.29 & 10.00 & .15 \\
6984 &  11.41 & 10.58 & 10.31 & 10.01 & .09 \\
7749 &  11.39 & 10.53 & 10.22 &  9.82 & .06 \\
7968 &  11.40 & 10.54 & 10.20 &  9.85 & .14 \\
8042 &  11.42 & 10.53 & 10.19 &  9.80 & .10 \\
8132 &  11.45 & 10.54 & 10.20 &  9.89 & .07 \\
8167 &  11.42 & 10.52 & 10.23 &  9.61 & .08 \\
8172 &        &       & 10.21 &  9.80 & .07 \\
8733 &  11.48 & 10.59 & 10.31 &  9.94 & .09 \\
8876 &  11.43 & 10.55 & 10.29 &  9.98 & .06 \\
9614 &  11.44 & 10.57 & 10.30 &  9.96 & .14 \\
9889 &  11.39 & 10.52 & 10.19 &  9.91 & .12 \\
9966 &  11.35 & 10.40 & 10.00 &  9.52 & .10 \\
10059&        &       & 10.15 &       &     \\
10213&  11.41 & 10.54 & 10.28 & 10.27 & .18 \\
10302&  11.41 & 10.55 & 10.31 & 10.10 & .13 \\
10614&  11.40 & 10.52 & 10.25 &  9.89 & .09 \\
10655&  11.34 & 10.45 & 10.07 &  9.64 & .08 \\
10742&  11.41 & 10.54 & 10.16 &  9.61 & .22 \\
10949&  11.39 & 10.51 & 10.14 &  9.62 & .06 \\
\end{tabular}                            
\end{table}                                   
                                  
\begin{table}                     
\caption{MKN509}                  
\begin{tabular}{lllll}           
MJD  &  $J$  &  $H$  &  $K$  &  $L$   \\
5235 &  11.98 & 11.07 & 10.20 &       \\    
5236 &  11.96 & 11.06 & 10.19 &       \\
5237 &  11.97 & 11.05 & 10.17 &       \\
5238 &  12.01 & 11.08 & 10.19 &       \\
5239 &  11.96 & 11.08 & 10.18 &       \\
5240 &  11.97 & 11.09 & 10.20 &       \\
5595 &  11.80 & 10.87 & 09.98 &  8.69 \\
6195 &  11.91 & 10.92 & 09.96 &  8.58 \\
6197 &  11.94 & 10.93 & 09.96 &       \\
6287 &  11.97 & 11.01 & 10.04 &  8.59 \\
6663 &  11.92 & 11.02 & 10.11 &  8.63 \\
6694 &  11.92 & 11.01 & 10.12 &  8.76 \\
6985 &  11.93 & 10.99 & 10.11 &  8.78 \\
7342 &  11.93 & 11.08 & 10.18 &  8.90 \\
7436 &  12.10 & 11.03 & 10.13 &  8.78 \\
7634 &  11.98 & 11.01 & 10.09 &       \\
7749 &  12.01 & 11.02 & 10.13 &  8.77 \\
7840 &  12.07 & 11.10 & 10.19 &  8.71 \\
7841 &  12.05 & 11.10 & 10.18 &  8.68 \\
8042 &  11.89 & 11.00 & 10.11 &  8.74 \\
8132 &  11.80 & 10.92 & 10.05 &  8.72 \\
8167 &  11.85 & 10.89 & 10.04 &  8.78 \\
8172 &  11.81 & 10.88 & 10.02 &  8.65 \\
8572 &  12.15 & 11.23 & 10.37 &  8.88 \\
8574 &  12.13 & 11.25 & 10.38 &  8.90 \\
8576 &  12.09 & 11.25 & 10.37 &  8.86 \\
8733 &  12.07 & 11.14 & 10.32 &  8.90 \\
8734 &  12.05 & 11.17 & 10.32 &  8.87 \\
8875 &  11.89 & 11.00 & 10.19 &  8.85 \\
8876 &  11.83 & 10.98 & 10.19 &  8.92 \\
8905 &  11.85 & 10.95 & 10.11 &       \\
9609 &  12.03 & 11.10 & 10.19 &  8.81 \\
9614 &  12.08 & 11.10 & 10.18 &  8.80 \\
9889 &  12.01 & 11.01 & 10.10 &  8.80 \\
9891 &  11.97 & 11.01 & 10.10 &  8.80 \\
9966 &  11.93 & 11.01 & 10.12 &  8.70 \\
10212&  11.94 & 11.04 & 10.14 &  8.68 \\
10394&  11.96 & 10.99 & 10.12 &  8.80 \\
10614&  12.24 & 11.27 & 10.37 &  8.97 \\
10655&  12.16 & 11.23 & 10.37 &  8.95 \\
10720&  12.06 & 11.19 & 10.33 &  8.89 \\
10749&  12.09 & 11.22 & 10.33 &       \\
10948&  12.01 & 11.09 & 10.25 &  8.81 \\
11117&  11.99 & 11.00 & 10.10 &  8.87 \\
\end{tabular}                                
\end{table}                                   
                              
\clearpage

\begin{table}                 
\caption{H2107-097}           
\begin{tabular}{llllll}       
MJD  &  $J$  &  $H$  &  $K$  &  $L$  & $L_{\rm err}$ \\
6696  & 12.38 & 11.42 & 10.54 & 9.36 & .09 \\
7841  & 12.35 & 11.34 & 10.51 & 9.26 & .08 \\
9971  & 12.40 & 11.44 & 10.55 & 9.14 & .07 \\
10304 & 12.43 & 11.45 & 10.61 & 9.25 & .07 \\
10656 & 12.51 & 11.64 & 10.81 & 9.46 & .06 \\
10657 & 12.60 & 11.62 & 10.82 &      &     \\
10948 & 12.64 & 11.61 & 10.84 & 9.45 & .03 \\
\end{tabular}                                
\end{table}                                   

\begin{table}
\caption{1H2129-624}
\begin{tabular}{llllll}
MJD  &  $J$  &  $H$  &  $K$  &  $L$  & $L_{\rm err}$ \\
6691  & 13.34 & 12.31 & 11.43 &  9.95 & .09 \\
8875  & 13.02 & 11.94 & 11.09 &  9.84 & .06 \\
9615  & 13.51 & 12.60 & 11.81 & 10.26 & .12 \\
10304 & 13.15 & 12.28 & 11.60 & 10.54 & .14 \\
10658 & 13.28 & 12.33 & 11.48 & 10.28 & .13 \\
10949 & 13.33 & 12.47 & 11.70 & 10.32 & .18\\
\end{tabular}                                      
\end{table}                                   

\begin{table}
\caption{NGC7213}
\begin{tabular}{lllll}
MJD  &  $J$  &  $H$  &  $K$  &  $L$   \\
3024  & 10.04  & 9.26  & 8.95  &      \\
4466  &  9.98  & 9.21  & 8.81  &      \\   
5304  & 10.03  & 9.28  & 8.86  & 8.12 \\
5305  & 10.05  & 9.29  & 8.85  & 8.15 \\
5890  & 10.02  & 9.20  & 8.83  & 8.17 \\
6197  & 10.06  & 9.24  & 8.87  & 8.20 \\
6287  & 10.00  & 9.20  & 8.83  & 8.15 \\
6582  & 10.02  & 9.20  & 8.80  & 8.11 \\
6692  & 10.04  & 9.23  & 8.85  & 8.18 \\
6780  & 10.07  & 9.23  & 8.84  & 8.18 \\
6984  & 10.04  & 9.24  & 8.87  & 8.12 \\
7342  & 10.05  & 9.25  & 8.85  & 8.10 \\
7437  & 10.05  & 9.22  & 8.85  & 8.15 \\
7749  & 10.03  & 9.21  & 8.82  & 8.06 \\
7843  & 10.00  & 9.18  & 8.79  & 7.99 \\
8173  & 10.06  & 9.18  & 8.79  & 7.99 \\
8529  & 10.07  & 9.22  & 8.87  & 8.16 \\
8575  & 10.07  & 9.25  & 8.88  & 8.17 \\
8874  & 10.05  & 9.23  & 8.88  & 8.23 \\
8905  & 10.08  & 9.25  & 8.92  & 8.27 \\
9614  & 10.07  & 9.27  & 8.94  & 8.25 \\
9890  & 10.09  & 9.27  & 8.96  & 8.45 \\
9970  & 10.08  & 9.26  & 8.95  & 8.27 \\
10213 & 10.07  & 9.24  & 8.95  & 8.32 \\
10304 & 10.10  & 9.27  & 8.97  & 8.39 \\
10658 & 10.06  & 9.26  & 8.93  & 8.38 \\
\end{tabular}                                    
\end{table}                                     

\begin{table}
\caption{NGC7469}
\begin{tabular}{lllll}
MJD  &  $J$  &  $H$  &  $K$  &  $L$   \\
5279  &       &        & 9.28  & 8.21 \\
5595  & 10.96 &  9.98  & 9.32  & 8.22 \\
5887  & 10.91 &  9.94  & 9.22  & 8.10 \\
5890  & 10.93 &  9.92  & 9.20  & 8.12 \\
6197  & 10.97 &  9.98  & 9.26  & 8.13 \\
6287  & 10.92 &  9.92  & 9.19  & 8.04 \\
6663  & 11.00 & 10.01  & 9.33  & 8.18 \\
6695  & 11.04 & 10.02  & 9.34  & 8.17 \\
6778  & 10.99 & 10.05  & 9.36  & 8.26 \\
6984  & 11.00 & 10.01  & 9.36  & 8.18 \\
7342  & 11.00 & 10.05  & 9.43  & 8.18 \\
7436  & 11.06 & 10.05  & 9.40  & 8.31 \\
7753  & 11.12 & 10.22  & 9.72  & 8.85 \\
7841  & 11.12 & 10.24  & 9.71  & 8.64 \\
8132  & 11.00 & 10.01  & 9.35  & 8.28 \\
8172  & 11.07 & 10.06  & 9.41  & 8.31 \\
8529  & 10.95 &  9.94  & 9.21  & 8.04 \\
8574  & 10.98 &  9.94  & 9.23  & 8.09 \\
8874  & 10.95 &  9.95  & 9.27  & 8.08 \\
8908  & 10.99 &  9.96  & 9.28  & 8.09 \\
9614  & 11.05 & 10.02  & 9.35  & 8.20 \\
9889  & 11.00 & 10.03  & 9.43  & 8.27 \\
9966  & 10.98 & 10.00  & 9.30  & 8.14 \\
10057 & 11.01 & 10.04  & 9.35  & 8.20 \\
10212 & 10.95 &  9.98  & 9.28  & 8.19 \\
10262 & 10.96 &  9.99  & 9.27  & 8.07 \\
10302 & 11.00 & 10.01  & 9.36  & 8.18 \\
10413 & 10.92 &  9.93  & 9.25  & 8.16 \\
10614 & 10.93 &  9.92  & 9.22  & 8.06 \\
10655 & 10.89 &  9.93  & 9.23  & 8.09 \\
10721 & 10.81 &  9.83  & 9.11  & 7.97 \\
10738 & 10.83 &  9.84  & 9.09  & 7.90 \\
11114 & 10.91 &  9.86  & 9.09  & 7.97 \\
\end{tabular}                                  
\end{table}                                   

\begin{table}
\caption{MCG-2-58-22}
\begin{tabular}{llllll}
MJD  &  $J$  &  $H$  &  $K$  &  $L$  & $L_{\rm err}$ \\
4484  & 12.45 & 11.45 & 10.54 &  9.10 & .09  \\  
4466  & 12.34 & 11.43 & 10.52 &       &      \\
5887  & 12.08 & 11.16 & 10.32 &  9.08 & .06  \\
5890  & 12.15 & 11.18 & 10.32 &  9.15 & .04  \\
6197  & 12.11 & 11.12 & 10.20 &  8.99 & .05  \\
6287  & 12.10 & 11.12 & 10.17 &  9.06 & .04  \\
6664  & 12.20 & 11.24 & 10.25 &  8.99 & .08  \\
\end{tabular}                                
\end{table}                                   

\begin{table}
\caption{NGC7603}
\begin{tabular}{llllll}
MJD  &  $J$  &  $H$  &  $K$  &  $L$  & $L_{\rm err}$ \\
3719  & 11.90 & 11.15 & 10.76 &       &     \\
5886  & 11.98 & 11.11 & 10.61 &       &     \\
6049  & 11.87 & 10.99 & 10.42 &       &     \\
6287  & 11.85 & 10.86 & 10.37 &  9.10 &  .05\\
6663  & 11.90 & 11.06 & 10.51 &       &     \\
6693  & 11.84 & 10.98 & 10.40 &  9.51 &  .20\\
6984  & 11.94 & 11.03 & 10.43 &  9.50 &  .08\\
7142  & 11.88 & 11.04 & 10.41 &  9.63 &  .22\\
7437  & 12.01 & 11.10 & 10.59 &  9.66 &  .12\\
7843  & 11.93 & 11.09 & 10.62 &  9.73 &  .13\\
8132  & 12.00 & 11.08 & 10.53 &  9.62 &  .04\\
8172  & 11.95 & 11.06 & 10.51 &  9.75 &  .06\\
8529  & 11.88 & 11.01 & 10.43 &  9.47 &  .07\\
8575  & 11.93 & 11.05 & 10.45 &  9.54 &  .10\\
8874  & 11.88 & 10.99 & 10.48 &  9.46 &  .04\\
9615  & 11.88 & 10.93 & 10.24 &  8.99 &  .04\\
9889  & 11.89 & 10.97 & 10.46 &  9.47 &  .09\\
9966  & 11.86 & 10.98 & 10.42 &  9.27 &  .08\\
10058 & 11.70 & 10.85 & 10.26 &  9.07 &  .07\\
10302 & 11.77 & 10.79 & 10.09 &  8.94 &  .05\\
10413 & 11.82 & 10.86 & 10.14 &  8.94 &  .05\\
10614 & 11.72 & 10.76 & 10.09 &  8.93 &  .05\\
10655 & 11.72 & 10.81 & 10.10 &  8.88 &  .05\\
10720 & 11.70 & 10.76 & 10.03 &  8.95 &  .07\\
10738 & 11.72 & 10.78 & 10.04 &  8.79 &  .10\\
11114 & 11.61 & 10.65 & 9.89  &  8.80 &  .05\\
\end{tabular}                                
\end{table}                                   
                                   

\begin{thebibliography}{99}

\bibitem[]{}Baribaud T., Alloin D., Glass, I.S., Pelat D., 1992, A\&A, 256, 375 
\bibitem[]{}Barvainis R., 1987, ApJ, 320, 537
\bibitem[]{}Barvainis R., 1992, ApJ, 400, 502
\bibitem[]{}Carter B.S., 1990, MNRAS, 242, 1
\bibitem[]{}Carone T.E. et al, 1996, ApJ, 471, 737  
\bibitem[]{}Cho{\l}oniewski J., 1981, Acta Astr, 31, 293
\bibitem[]{}Chuvaev K.K., Lyutyi V.M., Doroshenko V.T., 1990, Sov. Astr.
Letts., 16, 372
\bibitem[]{}Clavel J., Wamsteker W., Glass I.S., 1989, ApJ, 337, 236.
\bibitem[]{}Clements S.D., Smith A.G., Aller H.D., Aller M.F., 1995, AJ, 529
\bibitem[]{}Doroshenko V.T., 1996, Astr. Letts., 22, 309
\bibitem[]{}Doroshenko V.T., Lyuty V.M., 1999, Astr. Letts., 25, 883
\bibitem[]{}Doroshenko V.T., Lyutyi V.M., Rakhimov V.Yu., 1989, Sov. Astr.
Letts., 15, 207 
\bibitem[]{}Frogel J.A., Persson S.E., Aaronson M., Matthews K., 1978, ApJ,
220, 75 
\bibitem[]{}Gaskell C.M., Peterson B.M., 1987, ApJS, 65, 1
\bibitem[]{}Genzel R., Weitzel L., Tacconi-Garman L.E., Blietz M., Cameron
M., Krabbe A., Lutz D., Sternberg A., 1995, ApJ, 444, 129
\bibitem[]{}Glass I.S., 1984, MNRAS 211, 461 
\bibitem[]{}Glass I.S., 1986, MNRAS 219, 5p (Fairall 9)
\bibitem[]{}Glass I.S., 1992, MNRAS 256, L23. (NGC\,3783)
\bibitem[]{}Glass I.S., 1995, MNRAS 276, L65 (NGC\,1068)
\bibitem[]{}Glass I.S., 1997a, MNRAS 257, L50 (NGC\,2992)
\bibitem[]{}Glass I.S., 1997b, ApSpSci, 248, 191
\bibitem[]{}Glass I.S., 1997c, Mon Notes astr Soc S. Africa, 56, 110
\bibitem[]{}Glass I.S., 1998, MNRAS 297, 18 (NGC\,7469)
\bibitem[]{}Glass I.S., Carter B.S., 1989, In IAU Joint Commission  Meeting 
{\it Problems of Infrared Standardisation and Extinction}, ed. E.F. Milone,
Springer, Berlin, 37
\bibitem[]{}Griersmith D., Hyland A.R., Jones T.J., 1982, AJ, 87, 1106
\bibitem[]{}Hagen-Thorn V.A., Marchenko S.G., Mikolaichuk O.V., Yakoleva
V.A., 1997, Astr. Rep., 41, 1
\bibitem[]{}Kim C., Boller T., 2002, ApSpSci 281, 663
\bibitem[]{}Kollatschny W., Bischoff K., Dietrich M., 2000, A\&A, 361, 901
\bibitem[]{}Lebofsky M.J., Rieke G.H., 1980, Nature, 284, 410
\bibitem[]{}Longmore A.J., Sharples R.M., 1982, MNRAS, 201, 111
\bibitem[]{}Maiolino R., Marconi A., Salvati M., Risaliti G., Severgnini P.,
Oliva E., La Franca F., Vanzi L., 2001, A\&A, 365, 28
\bibitem[]{}Nelson B.O., 1996, ApJ 465, L87
\bibitem[]{}Oknyanskij V.L., 2002, in Galaxies, the Third Dimension, ASP
Conf.\ Ser.\ Vol.\ 282, p.\ 330
\bibitem[]{}Oknyanskij V.L., Horne K., 2001, in Probing the Physics of
Active Galactic Nuclei by Multiwavelength Monitoring, ASP Conf.\ Ser.\ Vol.\
224, p.\ 149 
\bibitem[]{}Penston M.V., Penston M.J., Selmes R.A., Becklin E.E.,
Neugebauer G., 1974, MNRAS, 169, 357
\bibitem[]{}Pier E.A. \& Krolik J.H., 1992, ApJ, 401, 99
\bibitem[]{}Pier E.A. \& Voit G.M., 1995, ApJ, 450, 628
\bibitem[]{}Salvati M. et al, (1993), A\&A, 274, 174
\bibitem[]{}Santos-Ll\'{e}o M., Clavel J., Barr P., Glass I.S., Pelat D.,
Peterson B.M., Reichert G., 1995, MNRAS 270, 580
\bibitem[]{}Schlegel D.J., Finkbeiner D.P., Davis M., 1998, ApJ, 500, 525 
\bibitem[]{}Schmid H.M., Appenzeller I., Burch U., 2003, A\&A, 404, 505
\bibitem[]{}Simon T., Drake S.A., 1989. ApJ, 346, 305
\bibitem[]{}V\'{e}ron-Cetty M.-P., V\'{e}ron P., 2000, ESO Scientific Report 
No.\ 19
\bibitem[]{}Winkler H., 1997, MNRAS, 292, 273
\bibitem[]{}Winkler H., Glass I.S., van Wyk F., Marang F., Spencer
Jones J.H., Buckley D.A.H. and Sekiguchi K., 1992. MNRAS, 257, 659

\end{thebibliography}
\end{document}